\documentclass[11pt]{article}
\usepackage[centertags]{amsmath}
\usepackage{amsfonts}
\usepackage{newlfont}
\usepackage{graphicx}
\usepackage{eurosym}
\usepackage{epstopdf}
\usepackage{amsmath,amssymb, xcolor}
\usepackage{caption}
\usepackage{subcaption}
\newtheorem{theorem}{Theorem}[section]
\newtheorem{proposition}{Proposition}[section]

\newtheorem{lemma}{Lemma}[section]

\newtheorem{rem}{Remark}[section]

\newcommand{\F}{\cal{F}}

\newcommand{\G}{\cal{G}}

\newcommand{\Fu}{{\cal{F}}_u}

\newcommand{\Ft}{{\cal{F}}_t}
\newcommand{\Gs}{{\cal{G}}_s}

\newcommand{\Gt}{{\cal{G}}_t}
\newcommand{\Ht}{{\cal{H}}_t}

\newcommand{\E}{\mathbf{E}}

\newcommand{\e}{\mathrm{e}}

\newcommand{\lam}{\lambda}

\def\cB{{\cal B}}

\def\cF{{\cal F}}

\begin{document}

\title{CVA in fractional and rough volatility models}

\author{E. Al\`os\footnote{Dept. of Economics and Business, University Pompeu Fabra, Barcelona, \texttt{elisa.alos@upf.edu}}, F. Antonelli\footnote{DISIM, University of L'Aquila, \texttt{fabio.antonelli@univaq.it}}, A. Ramponi\footnote{Dept. Economics and Finance, University of Roma - Tor Vergata, \texttt{alessandro.ramponi@uniroma2.it}}, S. Scarlatti\footnote{Dept. Economics and Finance, University of Roma - Tor Vergata, \texttt{sergio.scarlatti@uniroma2.it}}}

\maketitle

\begin{abstract}
In this work we present a general representation formula for the price of a vulnerable European option, and the related CVA in stochastic (either rough or not) volatility models for the underlying's price, when admitting correlation with the default event.  We specialize it for some volatility models and we provide price approximations, based on the representation formula. We study numerically their accuracy, comparing the results with Monte Carlo simulations, and we run a theoretical study of the error.
We also introduce a seminal study of roughness influence on the claim's price.

\medskip

\noindent
\textbf{Keywords}: Credit Value Adjustment; Vulnerable Options; Rough volatility models; Intensity approach.
\end{abstract}

\section{Introduction}

\label{intro}

In recent years, the interest in including correctly the effects due to default risks in the derivatives' evaluation has grown immensely, generating a whole new field in mathematical finance. As it is usually the case, the seminal papers work on simple market models  (see \cite{Lando98}, \cite{Kl96}  \cite{DS}, \cite{ZP07}) built the way towards the extension to more complex and realistic ones.
We refer the reader to \cite{BCB}, \cite{BR}, \cite {BJR}, \cite{BV18}, and \cite{BRH18}  for  a sample of the interest and width of the topic.

Surely, considering stochastic volatility for derivatives subject to credit risks is an important modeling feature, even more so under volatility roughness that better reflects the market behavior, as much of the recent literature shows, and we refer the reader to \cite{ALS}, \cite{ALe21}, \cite{BFG}, \cite {GJR} just to quote some.

Inevitably, including  credit risk features and stochastic volatility in the same model makes the picture more complex and harder to manage. Often, this complexity leaves Monte Carlo simulations as the only option to evaluate derivatives' prices in such a framework, paying a remarkable price in terms of computational times.

Therefore, extensive work was done to develop representation formulas that might lead to handier expressions and to finding alternative computational methods, as in \cite{A}, and more recently in \cite{ARS2} and  \cite{AARS21}.

Lately,  see for instance \cite{GRP}, deep learning techniques have been applied to evaluate a derivative depending on multiple risks, whose price is corrected by a family of adjustments that go under the acronym of XVA. This approach is based on the price characterization as solution of a  Backward Stochastic Differential Equation and the application of machine learning to those equations. Nevertheless, those algorithms  though highly efficient in the predictive phase, remain computationally costly in the learning phase  still based on Monte Carlo simulations. Moreover, the network architecture (e.g. the number of hidden layers and the number of neurons per layer), as well as the choice of the hyperparameters which characterize the implementation of the learning algorithm (the activation function, the minimization procedure, the mini-batch size, etc.) certainly require very careful programming work and skills.

Here, we focus our attention on vulnerable European options,
that are options  subject to some default event concerning the solvability of the issuer, and we provide a general representation formula for the price and the required Credit Value Adjustment (CVA) for stochastic (either rough or not) volatility market models, when correlations among the driving processes are present.

The underlying idea is inherited from the papers by Al\`os et al. (\cite{A}, \cite{AL21}), where the authors remark that those prices have to depend upon the integrated mean variance (or the zero-strike variance swap process) and the Black \& Scholes pricing function, deriving a price representation by applying either anticipative or non-anticipative Stochastic Calculus.  To extend that technique to a model including  also the adjustment coming from the default probabilities, we use the so called ``intensity approach" (\cite{BR}, \cite{BJR} just to quote some), that we assume to be a diffusion. Correlations  among all the processes, underlying, volatility and intensity, are admitted, making much harder to exploit the properties of the single processes (such as affinity). In particular, the intensity-underlying correlation is related to the so-called right/wrong-way risk.

Exploiting the fact that the vulnerable option final value is the same as in the default-free case, the explicit knowledge of the Black \& Scholes pricing function, and anticipating Stochastic Calculus, 
we are able to provide a general and rather manageable representation formula for the CVA (and consequently the adjusted price) in terms of  the correlations between the asset price and, respectively, the stochastic volatility and the intensity process. 
This representation turns out to be quite general, including at once both classical and rough stochastic volatility models. 

Further, on the basis of that representation, we develop a manageable approximation formula, which never employs the computationally costly  Monte Carlo simulations. 
To achieve this goal,  we first notice that the zero-strike variance swap process is the optional projection of the
integrated mean variance onto the market filtration, which allows developing the main term  in a non-anticipating fashion by classical It\^o formula. Finally, freezing some terms at the initial time, and using the well-known Clark-Ocone-Haussman formula, we are able to write an approximation that makes the covariation processes explicit.

When choosing the intensity in the class of the affine models and the volatility either a diffusion or a rough one, the formula becomes quite easily computable, as we show by selecting a CIR default  intensity and  a volatility following a Heston, SABR or Rough Bergomi model.
In those specialized models, the approximation takes the shape of a ``first-order" expansion with respect to the correlation parameters among the Brownian motions driving the processes dynamics.

Finally, we briefly discuss a theoretical error estimate procedure  and we test our method accuracy and efficiency by running a numerical study of the rBergomi price model coupled with a CIR default intensity model, taking the  corresponding Monte Carlo price as benchmark.  The relative errors turn out to be rather small, while there is a very remarkable gain in computational times dropping from hours down to few tens of seconds. We test our results by varying the values of the correlation parameters and the maturity, moreover  we perform a
 brief sensitivity analysis with respect to the Hurst parameter, to enucleate the roughness influence on the claim's price.
 
The paper is structured as follows. In the next two sections, we first introduce the theoretical framework for CVA evaluation and then we provide our general representation and approximation  formulas for a market model characterized by a stochastic (either rough or not) volatility. Section 4 specializes the approximation formula to the previously mentioned models, while section 5 is dedicated to a brief discussion of the error estimates. 
In the final section, restricting to the more interesting rBergomi model, we present the numerical analysis. We refer the reader to the appendix for a very short primer on Malliavin Calculus.

\section{CVA Evaluation of Defaultable European Claims}
\label{sec1}

In this section, we briefly describe the general framework concerning defaultable derivatives. 
Let  $[0,T]$  be  a finite time interval and $(\Omega, \F, P)$ a complete probability space  endowed with a filtration  $\{\Ft\}_{t\in [0,T]}$, augmented with the
$P-$null sets and made right continuous. We also assume that all the processes have a c\'adl\'ag version.

The market is  described by an interest rate $r_t$, and a process $X_t$, representing an asset's log-price, which may depend on multiple  stochastic factors.  The filtration  $\{\Ft\}_{\{t\in [0,T]\}}$ is rich enough to support all those processes, and we assume to be  in absence of arbitrage, with  $P$ a given risk neutral measure, selected by some criterion. We denote the discount factor by $B(t, s)= \e^{-\int_t^s r_udu}$.

In this market,  a defaultable European contingent claim paying $f(X_T)$ at maturity is traded, where $f$ is some function to be specified. During its lifetime the claim is subject to default and we denote by $\tau$ (not necessarily a stopping time w.r.t. the filtration $\Ft$) the default time. At default, the contingent's value may be partially recovered by the creditor and $Z_t$ denotes an $\Ft-$measurable bounded recovery process.

To properly evaluate this type of derivative, we need to include the information generated by the default time. We denote by $\Gt$ the progressively enlarged filtration, that makes $\tau$ a $\Gt-$stopping time, that is $\Gt=\Ft \lor \sigma(\{\tau\le t\})$. Hence, denoting by
$H_t = \mathbf 1_{\{\tau \le t\}}$ and by $
\Ht$ its natural filtration, we choose $\Gt=\Ft\lor\Ht$.

We make the fundamental assumption, known as the  H-hypothesis (see e.g. \cite{GJLR10} and \cite{G14} and the references therein), that

\medskip

\noindent
(H)\qquad \qquad\qquad \qquad  Every $\Ft-$martingale remains a $\Gt-$martingale.\hfill

\medskip \noindent
Under this assumption  $B(t,s)S_s, \, s\ge t$ remains  a $\Gs-$martingale under the unique extension of the risk neutral probability to the filtration $\Gs$.
(To keep notation light, we do not indicate  the probability we use for the  expectations, assuming to be  the  one corresponding to the filtration in use).

In this setting,  the price a defaultable claim, with final value  $f(X_T)$,  default time $\tau$ and recovery process $\{Z_t\}_t$ is given by
\begin{equation}
\label{eqvalue1}
c^d(t,T) = \E[B(t,T) f(X_T) 1_{\{\tau > T\}} + B(t,\tau) Z_{\tau} 1_{\{t < \tau \leq T\}} |\Gt],\quad t\in [0,T],
\end{equation}
while the corresponding default free  value  is
$
c(t,T) = \E[B(t,T)   f(X_T) |\Ft].
$

In many situations, investors do not know the default time and they may observe only whether it happened or not. The actual observable quantity is the asset price,  therefore  it  is interesting  to write the pricing formula \eqref{eqvalue1} in terms of $\Ft$, rather than in terms of $\Gt$. For that we have  the following key Lemma, see \cite{BJR} or \cite{BCB}.

\begin{lemma} For any integrable $\G-$measurable r.v. $Y$, the following equality holds
\begin{equation}
\label{key}
\E\Big[\mathbf 1_{\{\tau>
t\}}Y|\Gt\Big]=P(\tau>t|\Gt)\frac{\E\Big[\mathbf 1_{\{\tau>
t\}}Y|\Ft\Big]}{P(\tau>t|\Ft)}.
\end{equation}
\end{lemma}

Applying this lemma to the first and the second term of \eqref{eqvalue1}
and recalling that $1-H_t=\mathbf 1_{\{\tau>t\}}$ is $\Gt-$measurable, we obtain
\begin{eqnarray}
\label{key1}
\E[B(t,T) f(X_T) 1_{\{\tau > T\}}|\Gt]&=&\mathbf 1_{\{\tau>t\}}
\frac{\E[B(t,T) f(X_T) 1_{\{\tau > T\}}|\Ft]}{P(\tau>t|\Ft)},\\
\label{key1.2}
\E[B(t,\tau) Z_{\tau} 1_{\{t < \tau \leq T\}} |\Gt]&=&\mathbf 1_{\{\tau>t\}}\frac{\E[B(t,\tau) Z_{\tau} 1_{\{t < \tau \leq T\}} |\Ft]}{P(\tau>t|\Ft)},
\end{eqnarray}
which  may be made more explicit by following the hazard process  approach.

We denote  the conditional distribution of the default time  $\tau$ given $\Ft$ by
\begin{eqnarray}
F_t=P(\tau\leq t|\Ft), \qquad \forall \, t\ge 0,
\end{eqnarray}
whence, for $u\ge t$,  $P(\tau\leq u|\Ft)=\E(P(\tau\leq u|\Fu)|\Ft)=\E(F_u|\Ft)$. We also assume that $F_t(\omega)<1$ for all  $t>0$ to well define the so-called hazard process
\begin{equation}
\label{risk}
\Gamma_t:=-\ln(1-F_t)\quad\Rightarrow \quad F_t= 1 - \mathrm e^{-\Gamma_t}\quad \forall \, t>0, \qquad  \Gamma_0=0.
\end{equation}
With this notation, by an extension of Proposition 5.1.1 of \cite{BR}, we rewrite (\ref{key1}) and  \eqref{key1.2}  as
$$
\begin{aligned}
\E[B(t,T) f(X_T) 1_{\{\tau > T\}}|\Gt]=&\mathbf 1_{\{\tau>t\}}\E[B(t,T) f(X_T) \mathrm e^{-(\Gamma_T-\Gamma_t)}|\Ft],\\
\E[B(t,\tau) Z_{\tau} 1_{\{t < \tau \leq T\}} |\Gt]=&\mathbf 1_{\{\tau>t\}}\E\Big [\int_t^T B(t,s) Z_s\mathrm e^{-(\Gamma_s-\Gamma_t)}d\Gamma_s |\Ft\Big ].
\end{aligned}
$$
Assuming that  the hazard process  is differentiable with derivative $\lambda_t$, called the intensity process,
we arrive at the pricing formula
\begin{equation} \label{defaultprice}
\!c^d(t,T)\! = \!\mathbf 1_{\{\tau > t\}} \!\left[ \E\Big (\mathrm e^{-\!\int_t^T (r_s\!+\lambda_s) ds}f(X_T) \!+\!\!\int_t^T\!\!\! \!Z_s\lambda_s \mathrm e^{-\!\int_t^s (r_u\!+\lambda_u) du} ds| \Ft\Big ) \right]\! ,
\end{equation}
recovering  Lando's formulas (3.1) and (3.3) in \cite{Lando98}.

This formula can be specialized  further, assuming  fractional recovery (\cite{DS}),  $Z_t = R c(t,T)$  for some $0\le R< 1$, and using the Optional Projection Theorem  (see \cite{N}  Theorem 4.16) to obtain
\begin{equation}
\label{price2}
\begin{aligned}
c^d(t,T) = \mathbf 1_{\{\tau > t\}}\Big[R \E\Big(\mathrm e^{-\int_t^T r_udu} f(X_T) | \Ft \Big)
+ (1-R) \E\Big(\mathrm e^{-\int_t^T (r_u+\lambda_u) du} f(X_T) | \Ft \Big )\Big ],
\end{aligned}
\end{equation}
which was used also by Fard in \cite{Fard15} with $f(x)=(\mathrm e^x-K)^+$, and that can be interpreted as a convex combination of the default free price and the price with default.

As a consequence we have an expression also for the unilateral CVA, defined as difference between the default free price and the adjusted price
\begin{equation}
\label{cva0}
CVA(t) := \mathbf 1_{\{\tau > t\}} [c(t,T) - c^d(t,T)]
=\mathbf  1_{\{\tau > t\}} (1-R) \E\Big [\mathrm e^{-\int_t^T r_udu} f(X_T) (1-\mathrm e^{-\int_t^T \lambda_u du}) | \Ft\Big ].
\end{equation}
Of course, the computability of these expectations will depend heavily on the modeling choices one makes for $X$ and $\lambda$.

\section{A representation formula for the CVA in stochastic volatility models}
\label{sec 2}

 In this section, we consider  a family of stochastic volatility models. For the sake of simplicity,  we assume zero fractional recovery ($R=0$), and the risk-free spot rate, $r$, to be  a deterministic function of time. These are not restrictive assumptions, since the following discussion can be easily extended to consider $0<R<1$, and  increasing the dimensionality of the problem, also a stochastic interest rate might be included. We assume that the probability space contains at least a 3-dimensional standard $\cF_t-$adapted Brownian motion $\mathbf B_s=(B^1_s, B^2_s, B^3_s)$, representing the randomness sources of the market.

Indeed, our market model is described by an asset log-price  process, $X_t$, whose dynamics  under a given risk neutral measure  is   
\begin{equation}
\label{SDEsystem0}
dX_t =(r_t -\frac12 \sigma_t^2)dt + \sigma_t dB_t, \qquad B_t=\sqrt{1-\eta^2} B^1_t + \eta B^2_t  ,
\end{equation}
where the volatility process, $\sigma_t$, is assumed to be square-integrable and adapted to the filtration generated by $\{B^2_t\}_{t \geq 0}$. The parameter $\eta$ describes the correlation between the volatility and the log-price.

We remark that the above model includes the classical volatility models, where $\sigma$ is represented by a diffusion, but also those where the volatility's dynamics is determined by a fractional Brownian motion,  in order to better describe the long-term and short-term behavior of the implied volatility (see \cite{CR98}, and \cite{ALV07}). 

From now on, we take the shorter notation $\E_t[\cdot]$ to denote the conditional expectation with respect to  the filtration $\cF_t$, and we consider the CVA problem given by formula \eqref{cva0},  for $R=0$ and $f(x)=(\e^x-\e^\kappa)^+$, for some $\kappa\in \mathbb R$. Thus we need to evaluate on $\{\tau>t\}$
\begin{equation}
\label{start}
 \e^{- \int_t^T r_u du}\E_t\Big [ (1-\e^{-\int_t^T \lambda_u du}) (\e^{X_T}- \e^\kappa)^+\Big ], 
\end{equation}
which has to depend (as in the default free case) on  the integrated mean variance 
$$
v_t =\frac 1{T-t}\int_t^T\sigma_u^2du .
$$ 
For $0\le t\le s \le T$, we denote by
$\displaystyle N^t_T=\e^{-\int_t^T\lambda_udu }$ and by
$$
\cB(s,x,\zeta) = \e^x N(d_+(s,x,\zeta))-\e^{\kappa} N(d_-(s,x,\zeta)), \,\, \text{with }\, d_{\pm}(s,x,\zeta) \equiv d_{\pm}= \frac{x-\kappa \pm \frac{\zeta^2}2(T-s)}{\zeta \sqrt{T-s}}, 
$$
the Black \& Scholes European call pricing function, where $N$  is the standard normal distribution function. We recall that  $\cB$ verifies
\begin{eqnarray}
&&\label{BStime}\begin{cases}&\partial_t \cB(s,x,\zeta)+\frac {\zeta^2}2  (\partial_{xx}-\partial_x) \cB(s,x,\zeta) + r \partial_x\cB(s,x,\zeta)- r\cB(s,x,\zeta)=0\\
&  \cB(T,x,\zeta)=(\e^x-\e^\kappa)^+\end{cases},\\
&&\label{BSvol}\partial_\zeta \cB(s,x,\zeta)=\zeta(T-s)(\partial_{xx}-\partial_x) \cB(s,x,\zeta),\\
&&\partial_x \cB(s,x,\zeta) = \e^x N(d_+), \quad (\partial_{xx} -\partial_x) \cB(s,x,\zeta)   = \frac{\e^x}{\zeta \sqrt{T-s}} N'(d_+),\\
\label{third}
&&(\partial_{xxx} -\partial_{xx}) \cB(s,x,\zeta) = \frac{\e^x}{\sigma \sqrt{T-s}} N'(d_+) \left(1-\frac{d_+}{\zeta\sqrt{T-s}} \right), \\
\label{fourth}
&&(\partial_{xx} -\partial_x)^2 \cB(s,x,\zeta)   = \frac{\e^x N'(d_+)}{\zeta^3 (T-s)^{\frac 32}}\left [d_+^2 -d_+\zeta\sqrt{T-s}-1\right ].
\end{eqnarray}
Since the final condition $\cB(T,x,\zeta)=(\e^x-\e^\kappa)^+$ does not depend on $\zeta$, we may rewrite the
 risk-neutral expectation \eqref{start} as
\begin{equation}
\label{ev1}
\E_t\left[\e^{- \int_t^T r_u du}(1-N^t_T) \cB(T,X_T,v_T)\right] \mathbf 1_{\{\tau > t\}}.
\end{equation}
\begin{proposition}
\label{repres1}
Assuming the market  model \eqref{SDEsystem0}, we have
\begin{equation}
\begin{aligned}
\label{repr}
\!\!\!CVA(t)=& \E_t\left[\e^{- \int_t^T r_u du}(1-N^t_T) \cB(t,X_t,v_t)\right]\\
+&\frac 12\E_t\left[(1-N^t_T)\!\!\int_t^T \!\! \e^{- \int_t^s r_u du}\sigma_s\left( \partial_{xxx}\!- \!\partial_{xx}\right)\cB(s,X_s,v_s)\sum_{i=1}^2\!\int_s^T\!\!\! D_s^{B^i}\! \sigma_u^2du\,d\langle B,B^i\rangle_s\right]\\
-&\E_t\left[N^t_T\int_t^T\!\! \e^{- \int_t^s r_u du}\sigma_s\partial_x \cB (s,X_s,v_s)\sum_{i=1}^2 \int_s^T \!\!\!D_s^{B^i}\! \lambda_udu\, d\langle B,B^i\rangle_s\right ],
\end{aligned}
\end{equation}
where by $D_s^{B^i}$ we denoted the Malliavin derivative with respect to the $i-$th component of the Brownian motion $\mathbf B$.
\end{proposition}

\noindent
\begin{Proof} We denote $\displaystyle R_s^{-1}= \e^{- \int_0^s r_u du}  $ and $\displaystyle Y_t= \int_t^T \sigma^2_udu$.
We are going to apply the anticipating It\^o formula (see \cite{AL21}) to the process $ Y_t$ and the function
$$
F(s,x,y)= R_s^{-1} \cB\left (s,x, \sqrt{\frac y{T-s}} \right ).
$$
Exploiting relations \eqref{BStime} and \eqref{BSvol}, we have
$$\begin{aligned}
\partial_s F(s,x,y)=& R_s^{-1} \Big ( \partial_s\cB+ \frac{y}{2(T-s)} (\partial_{xx}-\partial_x)\cB-r_s\cB\Big )\cB\left (s,x, \sqrt{\frac y{T-s}} \right ),\\
\partial_x F(s,x,y)=&R_s^{-1}\partial_x \cB\left (s,x, \sqrt{\frac y{T-s}}\right ) , \quad \partial_{xx} F(s,x,y)= R_s^{-1}\cB\left (s,x, \sqrt{\frac y{T-s}} \right ),\\
\partial_y F(s,x,y)=& \frac 12 R_s^{-1}(\partial_{xx}-\partial_x)\cB\left(s,x, \sqrt{\frac y{T-s}} \right ), \\
\partial_{xy }F(s,x,y)=& \frac {R_s^{-1}}2 (\partial_{xxx}-\partial_{xx})\cB\left (s,x, \sqrt{\frac y{T-s}} \right ).
\end{aligned}
$$
Keeping in mind that $v_t=\sqrt{\frac {Y_t}{T-t}}$,  we obtain
$$
\begin{aligned}
CVA(t)=&\E_t\left[R_{[t,T]}^{-1} (1-N^t_T) \cB(T,X_T,v_T)\right] =\E_t\left[R_{[t,T]}^{-1}(1-N^t_T) \cB(t,X_t,v_t)\right]\\
+&\E_t\left[ (1-N^t_T) \int_t^T\frac{ \eta}2R_{[t,s]}^{-1}  \sigma_s(D^-Y)_s (\partial_{xxx}-\partial_{xx}) \cB(s,X_s,v_s)ds\right]\\
+&\E_t\left[ (1-N^t_T) \int_t^T\sqrt{1- \eta^2} R_{[t,s]}^{-1}  \sigma_s\partial_x\cB(s,X_s,v_s)dB^1_s\right]\\
+&\E_t\left[ (1-N^t_T) \int_t^T \eta R_{[t,s]}^{-1}  \sigma_s\partial_x\cB(s,X_s,v_s)dB^2_s\right],
\end{aligned}
$$
where $R_{[t,s]}^{-1} =R_s^{-1} R_t$, and
$\displaystyle (D^-Y)_s = D^{B^2}_s\left( \int_s^T \sigma^2_udu\right )$.

By using the duality relation with respect to the two independent Brownian motions for the last two integrals, we have 
$$ 
\begin{aligned}
CVA(t)=&\E_t\left[R_{[t,T]}^{-1}(1-N^t_T) \cB(t,X_t,v_t)\right]\\
+&\E_t\left[ (1-N^t_T) \int_t^T\frac{ \eta}2R_{[t,s]}^{-1}  \sigma_s(\partial_{xxx}-\partial_{xx}) \cB(s,X_s,v_s)\int_s^TD^{B^2}_s \sigma^2_udu\,ds\right]\\
+&\E_t\left[ \int_t^T D^{B^1}_s(1-N^t_T)R_{[t,s]}^{-1}  \sigma_s\partial_x\cB(s,X_s,v_s) \sqrt{1- \eta^2} ds\right]\\
+&\E_t\left[  \int_t^T D^{B^2}_s(1-N^t_T)R_{[t,s]}^{-1}  \sigma_s\partial_x\cB(s,X_s,v_s)\eta  ds\right].
\end{aligned}
$$
Observing that
$$
\begin{aligned}
D^{B^i}_s(1-N^t_T)=&- N^t_TD^{B^i}_s\left (- \int_t^T \lambda_u du\right)= N^t_T \int_s^T D^{B^i}_s\lambda_u du,\\
\eta D^{B^2}_s\left( \int_s^T \sigma^2_udu\right )=&\eta \int_s^TD^{B^2}_s \sigma^2_udu=
\sum_{i=1}^2 \int_s^TD^{B^i}_s \sigma^2_ud\langle B, B^i\rangle_s,
\end{aligned}
$$
because the process $\lambda $ is $\cF_t-$adapted and $D^{B^1} \sigma^2_s=0$ ($\sigma $ depends only on $B^2$), we finally  get to 
$$
\begin{aligned}
\!\!\!CVA(t)=& \E_t\left[\e^{- \int_t^T r_u du}(1-N^t_T) \cB(t,X_t,v_t)\right]\\
+&\frac 12\E_t\left[(1-N^t_T)\!\!\int_t^T \!\! \e^{- \int_t^s r_u du}\sigma_s\left( \partial_{xxx}\!- \!\partial_{xx}\right)\cB(s,X_s,v_s)\sum_{i=1}^2\!\int_s^T\!\!\! D_s^{B^i}\! \sigma_u^2du\,d\langle B,B^i\rangle_s\right]\\
+&\E_t\left[N^t_T\int_t^T\!\! \e^{- \int_t^s r_u du}\sigma_s\partial_x \cB (s,X_s,v_s)\sum_{i=1}^2 \int_s^T \!\!\!D_s^{B^i}\! \lambda_udu\, d\langle B,B^i\rangle_s\right ]. \hskip3cm \square
\end{aligned}
$$
\end{Proof}
\section{The approximation formula}

The above representation formula is very compact, but  clearly the expectations cannot be evaluated in closed form and  they need being approximated. 

Besides, it is evident  that the CVA is determined by the default intensity, but also by the correlations among  volatility, price and intensity, still partially hidden in the factors given by $\cB$ and its derivatives. In what follows we construct an approximation  that makes those contributions explicit, so that it might  be possible to fully parametrize them, in view of a sensitivity analysis.

We therefore need  to handle  the dynamics of $\lambda$ and $\sigma$ less implicitly and to approximate $\cB$ and its derivatives, when evaluated on the anticipating process $v_t$.
To this aim, we introduce the martingale $\displaystyle N^t_s= \E_s \left ( \e^{-\int_t^T \lambda_u du}\right )$ (with final value $N^t_T$), whose dynamics can be explicitly described resorting to the bond pricing theory, when choosing the intensity  $\lambda$ in the class of affine processes. Indeed, 
\begin{equation}
\label{bond0}
N^t_s= \E_s\Big ( \mathrm e^{-\int_t^T\lambda_udu } \Big)=\mathrm e^{-\int_t^s\lambda_udu }E_s\Big ( \mathrm e^{-\int_s^T\lambda_udu } \Big)= \mathrm e^{-\int_t^s\lambda_udu }\e^{\varphi(T-s)\lam_s+ \psi(T-s)},
\end{equation}
for some deterministic differentiable  functions $\varphi$ and $\psi$ of time to maturity, which implies that
$$
\begin{aligned}
dN^t_s=& -N^t_s\Big \{ [ \lam_s +\varphi'(T-s)\lam_s + \psi'(T-s) ]ds +\varphi(T-s)d \lam_s-\frac 12\varphi^2(T-s)d\langle\lam,\lam\rangle_s \Big \}\\
=&- N^t_s  \varphi(T-s)dM^\lam_s,
\end{aligned}
$$
where $M^\lam$ denotes the martingale part of the process $\lam$.

As for the anticipating process $v_t$, we introduce the martingale
$$
M_s=\E_s\left( \int_0^T \sigma^2_udu \right ),\quad \text{whence} \quad v^2_s=\frac 1{T-s} \left(M_T-\int_0^s \sigma^2_udu\right ),
$$ 
while the so called zero-strike variance swap process is given by
\begin{equation}
\label{swapvar}
\hat v^2_s=\frac 1 {T-s}\E_s\left [ \int_s^T\sigma_u^2du \right ]=\frac{1}{T-s} \left(M_s-\int_0^s \sigma^2_u du \right)\quad \Rightarrow\quad  \E_s(v^2_s)=\hat v^2_s.
\end{equation}
Assuming that $\sigma\in\mathbb D^{1,2}$, we may apply the  Clark-Ocone-Haussman formula to deduce the dynamics of $M$
$$
M_s= \E\left[\int_0^T\!\!\! \sigma^2_u du\right ]+\! \int_0^s\!\!\E_s\left [ D^{B^2}_s\left ( \int_0^T\!\!\!  \sigma^2_u du\right )\right ]dB^2_s= \E\left[\int_0^T\!\!\!  \sigma^2_u du\right ]+\! \int_0^s\!\!\E_s\left [ \int_s^T \!\!\! D^{B^2}_s \sigma^2_u du\right ]dB^2_s.
$$
We remark that this approach does not require the volatility to be a Markovian process and it can be applied to a rough volatility, if one obtains a manageable martingale representation. Indeed, in the next section we are going to provide an explicit CVA approximation in the case of classical Heston and SABR models, as well as for the Rough Bergomi volatility.

By this new notation, we may rewrite the representation formula \eqref{repr} as
$$
\begin{aligned}
\!\!\!CVA(t)=& \E_t\left[\e^{- \int_t^T r_u du}(1-N^t_T) \cB \left (t,X_t, \sqrt {\frac{M_T-\int_0^t \sigma^2_u du}{T-t}} \right )\right]\\
+&\frac 12\E_t\left[(1-N^t_T)\!\!\int_t^T \!\! \e^{- \int_t^s r_u du}\sigma_s\left( \partial_{xxx}\!- \!\partial_{xx}\right)\cB(s,X_s,v_s)\sum_{i=1}^2\!\int_s^T\!\!\! D_s^{B^i}\! \sigma_u^2du\,d\langle B,B^i\rangle_s\right]\\
+&\E_t\left[N^t_T\int_t^T\!\! \e^{- \int_t^s r_u du}\sigma_s\partial_x \cB (s,X_s,v_s)\sum_{i=1}^2 \int_s^T \!\!\!D_s^{B^i}\! \lambda_udu\, d\langle B,B^i\rangle_s\right ],
\end{aligned}
$$
and   we  suggest a handy approximation formula for the second and third term. By substituting $v^2_s$ with its conditional expectation $\hat v^2_s$ (so $v_s=\sqrt{v^2_s} \approx \sqrt{\E_s(v^2_s)}= \hat v_s$)  in the derivatives of $\cB$, and  freezing the state variables at their initial values, those terms may be approximated by 
$$
\begin{aligned}
&\frac 12\E_t\left[(1-N^t_T)\!\!\int_t^T \!\! \e^{- \int_t^s r_u du}\sigma_s\left( \partial_{xxx}\!- \!\partial_{xx}\right)\cB(s,X_s,v_s)\sum_{i=1}^2\!\int_s^T\!\!\! D_s^{B^i}\! \sigma_u^2du\,d\langle B,B^i\rangle_s\right]\\
&+\E_t\left[N^t_T\int_t^T\!\! \e^{- \int_t^s r_u du}\sigma_s\partial_x \cB (s,X_s,v_s)\sum_{i=1}^2 \int_s^T \!\!\!D_s^{B^i}\! \lambda_udu\, d\langle B,B^i\rangle_s\right ]\\
\approx &\frac 12 (1-N^t_t)\left( \partial_{xxx}\!- \!\partial_{xx}\right)\cB(t,X_t,\hat v_t)
\E_t\left[\int_t^T \!\! \e^{- \int_t^s r_u du}\sum_{i=1}^2\E_s\left (\int_s^T\!\!\! D_s^{B^i}\! \sigma_u^2du\right )\sigma_s d\langle B,B^i\rangle_s\right]\\
+&\partial_x \cB (t,X_t,v_t)\E_t\left[\int_t^T\!\! \e^{- \int_t^s r_u du}\sum_{i=1}^2\E_s\left (N^t_T \int_s^T \!\!\!D_s^{B^i}\! \lambda_udu\right )\sigma_s d\langle B,B^i\rangle_s\right ]\\
=&\frac{ (1-N^t_t)}2\left( \partial_{xxx}\!- \!\partial_{xx}\right)\cB(t,X_t,\hat v_t)
\E_t\left[\int_t^T \!\! \e^{- \int_t^s r_u du} d\langle M,X\rangle_s\right]\\
-&\partial_x \cB (t,X_t,\hat v_t)\E_t\left[\int_t^T\!\! \e^{- \int_t^s r_u du}d\langle N^t,X\rangle_s\right ],
\end{aligned}, 
$$
where we used the Clark-Ocone formula to represent the martingales $N^t$ and $M$
\begin{equation}
\label{martrep}
dM_s= \sum_{i=1}^3\E_s\left (\int_s^T D^{B^i}_s\sigma^2_u du\right)dB^i_s, \quad dN^t_s= - \sum_{i=1}^3\E_s\left (N^t_T\int_s^T D^{B^i}_s\lam_u du \right) dB^i_s
\end{equation}
and consequently their covariation processes with $X$.  Notice that the higher-order terms in this expansion include  the quadratic variation and the covariance  of the martingales $N,X,M$. As these quantities depend on some small parameters in most models (as we will see in the examples), these terms will be higher-order as functions of these small parameters. 
For the first term, we apply the classical It\^o formula to the process $M$ and the function $F(m)= \cB\left(t,x, \sqrt{\frac{m-\Sigma_t}{T-t}} \right)$, that, by virtue of \eqref{BSvol}, has derivatives
$$
\begin{aligned}
\partial_mF(m)=&\frac 12(\partial_{xx}- \partial_x)\cB\left(t,x, \sqrt{\frac{m-\Sigma_t}{T-t}} \right),\\
\partial_{mm}F(m)=&\frac 14(\partial_{xx}- \partial_x)^2\cB\left(t,x, \sqrt{\frac{m-\Sigma_t}{T-t}} \right).\\
\end{aligned}
$$
Recalling the expression of $\hat v_s$, the PDE \eqref{BStime}, and expanding the first term in \eqref{repr} by integration by parts, we get
$$
\begin{aligned}
\E_t\left[R_{[t,T]}^{-1}(1-N^t_T) \cB(t,X_t,v_t)\right]=&(1-N^t_t) \cB(t,X_t,\hat v_t)\\
+&\frac  18\E_t\left[\int_t^T R_{[t,s]}^{-1}(1-N^t_s)(\partial_{xx}- \partial_x)^2\cB(t,X_t, \hat v_s) d\langle M,M\rangle_s\right]\\
-&\frac  12\E_t\left[\int_t^TR_{[t,s]}^{-1}(\partial_{xx}- \partial_x)\cB(t,X_t, \hat v_s)d\langle N^t,M\rangle_s\right], 
\end{aligned}
$$
which we may approximate by freezing $N^t$ and $\hat v$ at the initial time. 

Summarizing, 
assuming  model (\ref{SDEsystem0}) and \eqref{swapvar}, the  approximated value adjustment of a vulnerable European call option  is given by
\begin{equation}
\label{cva1bis}\boxed{
\begin{aligned}
CVA(t)=&\E_t\left[\e^{-\int_t^T r_udu}(1-N^t_T) \cB(T,X_T,v_T)\right]\\
\approx&(1\!-\!N^t_t )\cB(t,X_t,\hat v_t)\\
&+\frac18 (1\!-\!N^t_t )(\partial_{xx}-\partial_{x})^2 \cB(t,X_t,\hat v_t) \E_t\Big [\!\int_t^T\!\!\! \!\e^{-\int_t^s r_udu}d\langle M\! ,M\rangle_s\Big ]\\
&+\frac12 (1\!-\!N^t_t )(\partial_{xxx}-\partial_{xx}) \cB(t,X_t,\hat v_t)\E_t\Big [\!\int_t^T\!\!\! \!\e^{-\int_t^s r_udu}d\langle M\! ,X\rangle_s\Big ]\\
&- \frac12  (\partial_{xx}-\partial_{x})\cB(t,X_t,\hat v_t) \E_t\Big [\!\int_t^T\!\!\! \e^{-\int_t^s r_udu}d\langle N^t\!,M\rangle_s\Big]\\
&-\partial_x \cB(t,X_t,\hat v_t)\E_t\Big [\!\int_t^T\!\!\! \!\e^{-\int_t^s r_udu}d\langle N^t\! ,X\rangle_s\Big ].
\end{aligned}}
\end{equation}
\begin{rem}In the above formula, we could leave the factor $(1-N^t_s)$ inside the integrals rather than freezing it also, at the initial time. Eventually, the two choices give about the same numerical contribution to the final results,  thus we opted for a handier expression, to compute the various terms in a fast and simple way.
\end{rem}

Expression  \eqref{cva1bis}  might be interpreted as a {\bf first-order approximation formula}, indeed when restricting to some specific models for $\lambda$ and $\sigma$, the final evaluation of the covariation processes will make the correlation parameters appear with the first power. Because of its integrability properties, in what follows, we take $\lambda$ as a CIR process verifying Feller's condition to ensure positivity, while for $\sigma$ we make some different choices, trying to understand also  the effect of roughness on the evaluation.
In the last section, we show numerically (comparing with Monte Carlo simulations) how accurate this approximation can be.

To simplify our discussion, from now on we assume $r\equiv 0$.

\section{The stochastic volatility-CIR intensity model}

\label{sec3}

Here,  we consider a CIR dynamics for the intensity process
\begin{equation}
\label{CIRint}
d\lam_s= q(\mu-\lam_s) ds + c \sqrt{\lam_s}dW_s, \qquad \lambda_t = \lambda>0, \quad s>t
\end{equation}
given that $c^2< 2q\mu$, with $c, \mu, q>0$ and 
\begin{equation}
W_s =\frac{\rho-\eta\gamma }{\sqrt{1-\eta^2}}B^1_s+ \gamma B^2_s+ \sqrt{\frac{1-(\eta^2+ \gamma ^2+\rho^2)+2\gamma \eta\rho}{1-\eta^2}} B^3_s,
\end{equation}
with parameters $(\gamma , \eta,\rho)$ such that
\begin{equation} \label{corr_bounds}
\gamma ^2 <1, \,\,\, \eta^2< 1,\,\,\rho^2<1, \quad  \gamma ^2 +\rho^2+\eta^2<1+2\gamma \eta\rho, 
\end{equation}
to obtain the correlations: 
$
\langle B, B^2\rangle_s=\eta s,  \langle B, W\rangle_s=\rho s, \langle B^2, W\rangle_s=\gamma  s$.

By Fourier inversion, we know that the martingale $\displaystyle N^t_s= \e^{-\int_t^s\lambda_u du} \e^{ \varphi(T-s)\lambda_s + \psi(T-s)}$ has an explicit expression (see \cite{LL}) with
$$
\varphi(T-s)=- \frac { 2( \e^{p(T-s)}-1)}{ p-q+(p+q)\e^{p(T-s)} },\qquad
\psi(T-s)=-\frac {2q\mu}{c^2}\ln \big [ \frac { 2p\e^{(p+q)(T-s)} }{p-q+(p+q)\e^{p(T-s)}}\Big ],
$$
with $p^2= q^2+2c^2$ and it  has dynamics $dN^t_s= -c \varphi(T-s) N^t_s\sqrt{\lambda_s}dW_s$. 

We are going to couple this intensity process with different choices for the stochastic volatility.

\subsection{Heston - CIR}

\label{subsec3.7}

The market model (see \cite{He}) is given by
\begin{equation}
\label{hesmod}
\begin{aligned}
dX_s=&-\frac 12 \sigma^2_s ds +\sigma_sd(\sqrt{1-\eta^2} B^1_s+  \eta B^2_s),\\
d \sigma^2_s = &k(\theta-\sigma^2_s) ds + \nu \sigma_s dB^2_s, \quad \sigma_t=\sigma>0,
\end{aligned}
\end{equation}
and we assume that Feller's condition is verified ($2 k\theta>\nu^2$) also for the volatility,  to ensure the process positivity. 

Consequently, for $r<u$, we find that the Malliavin derivative verifies
$$
D_s\sigma^2_u= \nu \sigma_r- \int_s^u kD_s \sigma^2_\xi d\xi+ \nu \int_s^u\frac 1{2\sigma_\xi}D_s \sigma^2_\xi dB^2_\xi,
$$
whence taking expectation, and substituting  the result in \eqref{martrep}, we obtain
\begin{equation}
\label{malderH}
\begin{aligned}
\E_s(D_s\sigma^2_u)=&\nu \sigma_s\e^{-k(u-s)}\\
dM_s=& \frac \nu k(1-\e^{-k(T-s)}) \sigma_sdB^2_s \qquad M_0=\theta(1-\e^{-kT})
\end{aligned}
\end{equation}
(see \cite{AE}), so that we have 
\begin{eqnarray*}
d\langle M,X\rangle_s&=&\eta  \nu  \frac{1-\e^{-k(T-s)}}{k}\sigma_s^2ds, \\
d\langle M,M\rangle_s&=& \frac{\nu^2}{k^2} \left [1-\e^{-k(T-s)}\right ]^2 \sigma_s^2ds,\\
d\langle N^t,X\rangle_s&=&- \rho c \varphi(T-s) N^t_s\sigma_s \sqrt{\lambda_s} ds,\\
d\langle N^t,M\rangle_s&=&-  \gamma  \frac {\nu c} k  \varphi(T-s)  \left[1-\e^{-k(T-s)}\right] N^t_s\sigma_s ,\sqrt{\lambda_s}ds
\end{eqnarray*}
which implies the approximation formula
\begin{equation}
\label{cvaHC}
\begin{aligned}
CVA(t) \approx & (1\!-\!N^t_t )\cB(t,X_t,\hat v_t)\\
+&\frac18 (1\!-\!N^t_t ) (\partial_{xx}-\partial_{x})^2 \cB(tX_t,\hat v_t) \frac{\nu^2}{k^2} \int_t^T  \left[1-\e^{-k(T-s)}\right]^2 \E_t\left [\sigma_s^2 \right ] ds\\
+&\frac\eta2 (1\!-\!N^t_t ) (\partial_{xxx}-\partial_{xx}) \cB(t,X_t,\hat v_t) \frac{\nu}{k} \int_t^T  \left[1-\e^{-k(T-s}\right] \E_t\left[\sigma^2_s \right ]  ds \\
+& \frac\gamma 2  (\partial_{xx}-\partial_{x})\cB(t,X_t,\hat v_t)   \frac{\nu c}{k} \int_t^T \varphi(T-s)   \left[1-\e^{-k(T-s)}\right]\E_t \left[N^t_s \sqrt{\lambda_s}\sigma_s \right] ds \\
+& \rho c \partial_x \cB(t,X_t,\hat v_t) \int_t^T \varphi(T-s)  \E_t \left[ N^t_s\sqrt{\lambda_s} \sigma_s\right] ds.
\end{aligned}
\end{equation}
Recalling that $\E_t\left[\sigma^2_s \right ] =\theta +( \sigma_t^2-\theta)\e^{-k(s-t)}$, we can compute exactly the first two integrals. For the others, we remark that empirical experience from the market shows  that the dependence between default and stochastic volatility is rather weak, so that practitioners often consider them as independent. We assume this viewpoint to approximate
$\E_t (N^t_s\sqrt{\lambda_s}\sigma_s )$ by $\E_t( N^t_s\sqrt{\lambda_s})\E_t(\sigma_s)$, 
even if $\gamma \neq 0$. 

As in \cite{AARS21}, to compute $\E_t(N^t_s \sqrt{\lambda_s })$ we use the integration by parts  formula, and we write
\begin{equation}
\label{lamN}
d(\sqrt{\lambda_s}N^t_s )=\frac 12 \sqrt{\lambda_s}N^t_s\Big [ \frac {4q\mu - c^2}{4\lambda_s}- ( q+ c^2 \varphi(T\!-\!s))\Big ] ds +c N^t_s\Big [\frac 12-  \varphi(T-\!s)\lambda_s\Big ]dW_s,
\end{equation}
whence, by considering that the martingale part gives null contribution, we have
$$
\E_t(\sqrt{\lambda_s}N^t_s )=\sqrt \lambda N^t_t+ \int_t^s \frac 12\E_t\Big (\sqrt{\lambda_u}N^t_u\Big [ \frac {4q\mu - c^2}{4\lambda_u}- (q+ c^2 \varphi(T-u) )\Big ] \Big ) du.
$$
To approximate this last expectation, we freeze the $\frac 1{\lambda_s}$ factor at the initial value  and we solve the resulting ordinary differential  equation, obtaining
\begin{equation}
\label{Nsqlam}
\E_t(\sqrt{\lambda_s}N^t_s ) \approx \sqrt \lambda N^t_t \e^{\int_t^s g(u) du}, \quad
\textrm{ where}\quad
g(u)=\frac{4 q \mu -c^2}{8 \lambda} - \frac12(q  + c^2  \varphi(T-u)).
\end{equation}
Finally, the factor  $\E_t\Big [\sigma_s\Big ]$ can be computed by using a log-normal (moment-matching) approximation  as in \cite{AARS21}.

\subsubsection{SABR - CIR}

\label{subsec3.5}

We now consider the SABR model (\cite{Ha}). For $0\le \beta\le 1$ and $\alpha\in \mathbb R$, we have
\begin{eqnarray}
dX_s & = & -\frac{\sigma_s^2}{2} \e^{-2(1-\beta) X_s} ds + \sigma_s \e^{-(1-\beta) X_s} d(\sqrt{1-\eta^2} B^1_s+  \eta B^2_s), \\
d \sigma_s & = & \alpha \sigma_s dB^2_s.
\end{eqnarray}
In this case, for any $r<u$
$$
\sigma_u= \sigma_s \e^{\alpha( B^2_u-B^2_s) -\frac { \alpha^2}2 (u-s)}\quad \Rightarrow \quad \sigma^2_s \e^{2\alpha (B^2_u- B^2_s) -2 \alpha^2 (u-s)}
$$
whence
$$
\begin{aligned}
D_s\sigma^2_u=&2 \alpha \sigma^2_u \quad \Rightarrow \quad \E_s(D_s\sigma^2_u)=2\alpha \sigma^2_s\e^{\alpha^2(u-s)}, \quad \text{and}\\
dM_s=& 2\alpha \sigma_s^2\int_s^T \e^{\alpha^2(u-s)} du dB^2_s = 2 \sigma_s^2\frac{\e^{\alpha^2(T-s)} -1}{ \alpha}  dB^2_s, 
\end{aligned}
$$
and
\begin{equation}
\label{covar2}
\begin{aligned}
d\langle M, X\rangle_s    &= \eta \sigma_s^3 \e^{-(1-\beta)X_s}\frac{\e^{\alpha^2(T-s)} -1}{\alpha} ds, \\
d\langle M, M \rangle_s   &= 4\sigma_s^4 \frac{(\e^{2\alpha^2(T-s)} -1)^2}{\alpha^2} ds, \\
d\langle N^t, X\rangle_s &= - \rho c \varphi(T-s) N^t_s \sqrt {\lambda_s} \sigma_s \mathrm e^{-(1-\beta)X_s} ds,\\
d\langle N^t, M\rangle_s &= -2\gamma  c  \varphi(T-s)\frac{\e^{\alpha^2(T-s)} -1}{ \alpha} N^t_s \sqrt {\lambda_s}   \sigma_s^2 ds,
\end{aligned}
\end{equation}
so that  \eqref{cva1bis} may be approximated by
\begin{equation}
\label{CIRSABR1}
\begin{aligned}
CVA(t) \approx & (1\!-\!N^t_t )\cB(t,X_t,\hat v_t)\\
+&\frac12 (1\!-\!N^t_t )(\partial_{xx}-\partial_{x})^2 \cB(t,X_t,\hat v_t) \int_t^T\!\!\frac{(\e^{2\alpha^2(T-s)} -1)^2}{\alpha^2} \E_t\big [ \sigma_s^4 \big]ds\\
+&\frac \eta2(1\!-\!N^t_t )(\partial_{xxx}\!\!-\partial_{xx}) \cB(t,X_t,\hat v_t)\!\int_t^T\!\!\frac{\e^{\alpha^2(T\!-s)} \!-1}{\alpha} E_t\Big [ \sigma_s^3 \e^{-(1\!-\beta)X_s}\Big ]ds\\
+&\gamma  c   (\partial_{xx}-\partial_{x})\cB(t,X_t,\hat v_t) \E_t\Big [\!\int_t^T\!\!\! \varphi(T-s)\frac{\e^{\alpha^2(T-s)} -1}{ \alpha} N^t_s \sqrt {\lambda_s}   \sigma_s^2 ds\Big]\\
+&\rho c
\partial_x \cB(t,X_t,\hat v_t)\E_t\Big [\!\int_t^T\!\!\! \varphi(T-s) N^t_s \sqrt {\lambda_s} \sigma_s \mathrm e^{-(1-\beta)X_s} ds\Big ].
\end{aligned}
\end{equation}
To get to an actually implementable formula,
\begin{itemize}
\item we freeze the term $ \e^{-(1-\beta)X_s}$ at the initial condition, obtaining $ \e^{-(1-\beta)X_t}$;
\item for $n=1,2$, as before we treat the terms $\E_t\Big [ N^t_s \sqrt {\lambda_s}   \sigma_s^n\Big ]$ as regarding independent processes, obtaining
$\E_t\Big [ N^t_s \sqrt {\lambda_s} \Big ]\E_t\big [  \sigma_s^n\big]$;

\item we approximate $\E_t(N^t_s \sqrt{\lambda_s })$ as in \eqref{Nsqlam};
\item for $n=1,2,3,4$, we finally recall that
$\E_t(\sigma^n_s)=\sigma_t^n \e^{n(n-1) \frac {\alpha^2}2(s-t)}$.
\end{itemize}

\subsubsection{rBergomi - CIR}

In this subsection, we analyze our formula in the case of the Rough Bergomi volatility model.

In this case, the market model is still \eqref{SDEsystem0}, where $\sigma_s $ follows
\begin{equation}
\sigma_s^2=\sigma_0^2\e^{\nu\sqrt{2H}Z_s-\frac{\nu^2}2 s^{2H}},\quad \nu, \sigma_0>0, \quad H<\frac12,
\end{equation}
 with $Z$  a RLfBm of the form
$\displaystyle
Z_s:=\int_0^s (s-\xi)^{H-\frac12}dB^2_\xi.
$

Consequently
\begin{equation}
\begin{aligned}
D_s\sigma_u^2=&\nu\sqrt{2H}\sigma_u^2 D_s\left[     \int_0^u(u-\xi)^{H-\frac12}dB^2_\xi-\frac12 \nu^2u^{2H}               \right]=\nu\sqrt{2H}\sigma_u^2 D_s\left[    \int_s^u(u-\xi)^{H-\frac12}dB^2_\xi \right]\\
=&\nu\sqrt{2H}\sigma_u^2 (u-s)^{H-\frac12},\\
\Rightarrow &\qquad \E_s(D_s\sigma_u^2)=\nu\sqrt{2H} (u-s)^{H-\frac12}\E_s(\sigma_u^2).
\end{aligned}
\end{equation}
It follows that
\begin{equation}
dM_s=\nu\sqrt{2H}\int_s^T(u-s)^{H-\frac12} E_s\big(\sigma_u^2 \big)du\, dB^2_s,
\end{equation}
and therefore
\begin{equation}
\label{covarBer}
\begin{aligned}
d\langle M, X\rangle_s    &= \eta \nu\sqrt{2H} \sigma_s\int_s^T (u-s)^{H-\frac12}\E_s\big(\sigma_u^2 \big)du ds, \\
d\langle M, M \rangle_s   &= \nu^2  2H\left [\int_s^T (u-s)^{H-\frac12}\E_s\big(\sigma_u^2 \big)du\right ]^2ds, \\
d\langle N^t, X\rangle_s &= - \rho c \varphi(T-s) N^t_s \sqrt {\lambda_s} \sigma_s ds,\\
d\langle N^t, M\rangle_s &= -\gamma  c  \nu\sqrt{2H} \varphi(T-s)N^t_s \sqrt {\lambda_s}  \int_s^T (u-s)^{H-\frac12}\E_s\big(\sigma_u^2 \big)du  ds.
\end{aligned}
\end{equation}
When substituting in the approximation formula \eqref{cva1bis}, using the stochastic Fubini theorem, we finally obtain
\begin{equation}
\label{cva_rB}
\begin{aligned}
CVA(t)
\approx&(1\!-\!N^t_t )\cB(t,X_t,\hat v_t)\\
+&\frac{ \nu^2 H}4 (1\!-\!N^t_t )(\partial_{xx}\!\!-\partial_{x})^2 \cB(t,X_t,\hat v_t) \!\int_t^T\!\!\! \E_t\left [\Big [\int_s^T (u-s)^{H-\frac12}\E_s\big(\sigma_u^2 \big)du\Big ]^2\right ]ds\\
+&\frac\eta 2 \nu\sqrt{2H}  (1\!-\!N^t_t )(\partial_{xxx}\!\!-\partial_{xx}) \cB(t,X_t,\hat v_t)\!\int_t^T\!\!\! \int_s^T (u-s)^{H-\frac12}\E_t\Big [\sigma_s\E_s\big(\sigma_u^2 \big)\Big ]du ds\\
+&\rho \frac c 2  (\partial_{xx}\!\!-\partial_{x})\cB(t,X_t,\hat v_t) \!\int_t^T\!\!\!\varphi(T-s) \E_t\Big [N^t_s \sqrt {\lambda_s} \sigma_s \Big]ds\\
+&\gamma  c  \nu\sqrt{2H} \partial_x \cB(t,X_t,\hat v_t)\int_t^T\!\!\!\varphi(T-s) \int_s^T (u-s)^{H-\frac12}\E_t\Big [N^t_s \sqrt {\lambda_s} \E_s\big(\sigma_u^2 \big)\Big ]du  ds.
\end{aligned}
\end{equation}
To make the above expression more manageable, further computations  and approximations are required. 
\begin{itemize}
\item We first remark that 
$$
\begin{aligned}
\E_t[\sigma^2_u] =& \E_t\big (\sigma_0^2\e^{\nu\sqrt{2H}Z_u-\frac{\nu^2}2 u^{2H}}\big)=
\sigma_0^2\e^{-\frac{\nu^2}2 u^{2H}} \e^{\nu\sqrt{2H}\int_0^t (u-\xi)^{H-\frac12}dB^2_\xi}
E_t\big (\e^{\nu\sqrt{2H}\int_t^u (u-\xi)^{H-\frac12}dB^2_\xi}\big)\\
=&\sigma_0^2\e^{\frac{\nu^2}2 [(u-t)^{2H}-u^{2H}]} \,\e^{\nu\sqrt{2H}\int_0^t (u-\xi)^{H-\frac12}dB^2_\xi},
\end{aligned}
$$
since $\displaystyle \sqrt{2H}\int_t^u (u-\xi)^{H-\frac12}dB^2_\xi\sim N\big(0,(u-t)^{2H} \big)$. In the second term  we have $t\le s\le u$, so  $\E_t (\E_s(\sigma^2_u))=\E_t (\sigma^2_u)$.
\item It follows that in the first integral term we have 
$$
\begin{aligned}
\E_t\big(\sigma_s\E_s\big(\sigma_u^2 \big)\big)=&\E_t\Big(\sigma_s\sigma_0^2\e^{\frac{\nu^2}2 [(u-s)^{2H}-u^{2H}]} \,\e^{\nu\sqrt{2H}\int_0^s (u-\xi)^{H-\frac12}dB^2_\xi}\Big)\\
=& \E_t\Big(\sigma_0^3\e^{\frac{\nu^2}2 [(u-s)^{2H}-u^{2H}- \frac 12s^{2H}]} \,\e^{\nu\sqrt{2H}\int_0^s [(u-\xi)^{H-\frac12}+\frac 12 (s-\xi)^{H-\frac12}] dB^2_\xi}\Big)\\
=&\sigma_0^3\e^{\frac{\nu^2}2 [(u-s)^{2H}-u^{2H}-  \frac 12 s^{2H}]} \e^{\nu\sqrt{2H}\int_0^t [(u-\xi)^{H-\frac12}+\frac 12 (s-\xi)^{H-\frac12}] dB^2_\xi} \\
\times&\E_t\Big(\e^{\nu\sqrt{2H}\int_t^s [(u-\xi)^{H-\frac12}+\frac 12 (s-\xi)^{H-\frac12}] dB^2_\xi}\Big)\\
=&\sigma_0^3\e^{\frac{\nu^2}2 [(u-s)^{2H}-u^{2H}- \frac 12s^{2H}]} \e^{\nu^2\sqrt{2H}\int_0^t [(u-\xi)^{H-\frac12}+ \frac 12 (s-\xi)^{H-\frac12}] dB^2_\xi}\e^{\nu^2 H\Theta^2_s},
\end{aligned}
$$
where
\begin{equation} \label{csi}
\begin{aligned}
\Theta^2_s=&\int_t^s\big [(u-\xi)^{H-\frac12}+ \frac 12(s-\xi)^{H-\frac12}\big]^2 d\xi= \frac{(s-t)^{2H}}{4H} + \left(\frac{(u-t)^{2H}-(u-s)^{2H}}{2H} \right) \\
+&2 \frac{(s-t)^{H+\frac12} (u-t)^{H-\frac12}}{2H+1}  \left(\frac{t-u}{s-u} \right)^{\frac12-H} \!\!\!
\,_2F_{1}\Big(\frac12-H,H+\frac12;H+\frac32;\frac{s-t}{s-u}\Big),
\end{aligned}
\end{equation}
with $\,_{2}F_{1}$ the hypergeometric function\footnote{Symbolic computation by means of Wolfram Mathematica 12.2. and numerically tested on MatLab R2019b.}. 

\item
We observe that the expectation in the second integral term can be written as
$$
\label{dMM}
\begin{aligned}
&\E_t\left [\Big [\int_s^T\!\!\! (u\!-s)^{H-\frac12}\E_s\big(\sigma_u^2 \big)du\Big ]^2\right ]\\
=& \E_t\left[\int_s^T \!\!(u\!-\!s)^{H-\frac12}\E_s\big(\sigma_u^2 \big)du\!\! \int_s^T\!\!\!(\vartheta\!-\!s)^{H-\frac12} \E_s(\sigma_\vartheta^2)  d\vartheta \right]  \\
=&\int_s^T\!\!\! \int_s^T\!\! \sigma_0^4  \e^{\frac{\nu^2}2 [ (u-s)^{2H}-u^{2H}+(\vartheta\!-\!s)^{2H}-\vartheta^{2H}]}(u\!-\!s)^{H-\frac12}   (\vartheta\!-\!s)^{H-\frac12}\\
&\qquad \times\E_t\Big [ \e^{\nu \sqrt{2H} \int_0^s[(u-\xi)^{H-1/2} + (\vartheta-\xi)^{H-1/2}]dB^2_\xi} \Big ]du d\vartheta \\
= &
\sigma_0^4 \e^{\nu \sqrt{2H} \int_0^t[(u-\xi)^{H-1/2} + (\vartheta-\xi)^{H-1/2}]dB^2_\xi}\\
&\times \int_s^T\!\! \int_s^T  \!\!  \e^{\frac{\nu^2}2 [ (u-s)^{2H}-u^{2H}+(\vartheta\!-\!s)^{2H}-\vartheta^{2H}]}(u\!-\!s)^{H-\frac12}   (\vartheta\!-\!s)^{H-\frac12}  \e^{\nu^2 H \Phi^2_s}du d\vartheta ,
\end{aligned}
$$
where we used again the expression of the volatility second moment and we set\footnote{Symbolic computation by means of Wolfram Mathematica 12.2. and numerically tested on MatLab R2019b.}
\begin{equation} \label{zita}
\begin{aligned}
\Phi^2_s = &\int_t^s[(u\!-\!\xi)^{H-\frac12}+(\vartheta\!-\!\xi)^{H-\frac12}]^2 d\xi  = \frac{(u\!-\!t)^{2H}-(u\!-\!s)^{2H}+(\vartheta\!-\!t)^{2H}-(\vartheta\!-\!s)^{2H}}{2H} \\
+&2 \frac{\Gamma(H+\frac12)}{\Gamma(H+\frac32)(u-\vartheta)}\Big [(u-s)^{H+\frac12} (\vartheta-s)^{H+\frac12} {}_{2}F_{1}\Big(1,1+2H;H+\frac32;\frac{u-s}{u-\vartheta}\Big) \\
 - & (u-t)^{H+\frac12} (\vartheta-t)^{H+\frac12} \!\,_{2}F_{1}\Big(1,1+2H;H+\frac32;\frac{u-t}{u-\vartheta}\Big )\Big].
 \end{aligned}
\end{equation}

\item
It remains to compute the two expectations
$$
\E_t\Big [N^t_s \sqrt {\lambda_s} \sigma_s \Big], \qquad\E_t\Big [N^t_s \sqrt {\lambda_s} \E_s\big(\sigma_u^2 \big)\Big ].
$$
As before, we consider the volatility and the intensity processes as independent, hence
$$
\E_t\Big [N^t_s \sqrt {\lambda_s} \sigma_s \Big]=\E_t\Big [N^t_s \sqrt {\lambda_s}\Big] \E_t\big [\sigma_s \big], \quad\E_t\Big [N^t_s \sqrt {\lambda_s} \E_s\big(\sigma_u^2 \big)\Big ]=\E_t\Big [N^t_s \sqrt {\lambda_s}\Big] 
 \E_t\big [\sigma_u^2 \big ].
$$
We approximate $\E_t\Big [N^t_s \sqrt {\lambda_s}\Big] $ as in \eqref{Nsqlam}, $ \E_t\big [\sigma_u^2 \big ]$ was computed before, and similarly
$$
\begin{aligned}
\E_t\big [\sigma_s \big]=&\sigma_0\e^{\frac {\nu\sqrt{H} }{\sqrt 2}\int_0^t (s-\xi)^{H-\frac12}dB^2_\xi-\frac{\nu^2}4 
s^{2H}}\E_t\Big [\e^{\frac {\nu\sqrt{H} }{\sqrt 2}\int_t^s (s-\xi)^{H-\frac12}dB^2_\xi}\Big ]\\
=&\sigma_0\e^{\frac {\nu\sqrt{H} }{\sqrt 2}\int_t^s (s-\xi)^{H-\frac12}dB^2_\xi+
\frac {\nu^2 }4[ (s-t)^{2H}-s^{2H}] }
\end{aligned}
$$
\end{itemize}

\begin{rem} \label{disentangling}
By grouping all the terms, we finally get an approximation of the form
$$
\widehat{CVA}(t) = (1-N_t^t) c^{rf}_{rBergomi}(t,X_t,\hat v_t) + \rho g_1(t) + \gamma g_2(t),
$$ 
where $1-N_t^t$ is the survival probability of the counterparty, $c^{rf}_{rBergomi}$ is the price of the risk-free call under the rBergomi model and $g_{1,2}(t)$ are the respective contributions due to underlying-intensity and volatility-intensity correlations.
\end{rem}

\begin{rem} \label{nonmarkov}
From the previous computations, it appears clear that, by the Markov property, in the first two cases $CVA(t)$ is a deterministic function of the state variables, while the same is not verified in the case of the rough volatility models, indeed the conditional expectations, when computed, leave a stochastic integral term.
\end{rem}

\section{Error Estimates}

In this section, we want to estimate the error bounds  theoretically.We take $r\equiv 0$ for the sake of exposition. Starting from representation  We know from Proposition \ref{repres1} that 
\begin{equation}
\begin{aligned}
\label{repr}
\!\!\!CVA(t)=& \E_t\left[(1-N^t_T) \cB(t,X_t,v_t)\right]\\
+&\frac 12\E_t\left[(1-N^t_T)\!\!\int_t^T \!\sigma_s\left( \partial_{xxx}\!- \!\partial_{xx}\right)\cB(s,X_s,v_s)\sum_{i=1}^2\!\int_s^T\!\!\! D_s^{B^i}\! \sigma_u^2du\,d\langle B,B^i\rangle_s\right]\\
-&\E_t\left[N^t_T\int_t^T\!\sigma_s\partial_x \cB (s,X_s,v_s)\sum_{i=1}^2 \int_s^T \!\!\!D_s^{B^i}\! \lambda_udu\, d\langle B,B^i\rangle_s\right ]\\
=&T_1+T_2+T_3.
\end{aligned}
\end{equation}
Applying again the anticipating It\^o formula to the process
$$
(1-N^t_T)\left( \partial_{xxx}\!- \!\partial_{xx}\right)\cB(\theta,X_\theta,v_\theta)\!\!\int_\theta^T \!\! \sigma_s\sum_{i=1}^2\!\int_s^T\!\!\! D_s^{B^i}\! \sigma_u^2du\,d\langle B,B^i\rangle_s
$$
after taking expectations, keeping in mind that this process is $0$ when $\theta=T$, we obtain
\begin{equation}
\label{primera}
\begin{aligned}
0=&\E_t \left[ (1-N^t_T)\left( \partial_{xxx}\!- \!\partial_{xx}\right)\cB(t,X_t,v_t)\!\!\int_t^T \!\! \sum_{i=1}^2\!\int_s^T\!\!\! \sigma_sD_s^{B^i}\! \sigma_u^2du\,d\langle B,B^i\rangle_s  \right.\\
-&\left.(1-N^t_T)\!\!\int_t^T \!\! \e^{- \int_t^s r_u du}\sigma_s\left( \partial_{xxx}\!- \!\partial_{xx}\right)\cB(s,X_s,v_s)\sum_{i=1}^2\!\int_s^T\!\!\! D_s^{B^i}\! \sigma_u^2du\,d\langle B,B^i\rangle_s\right]+T_4,
\end{aligned}
\end{equation}
where $T_4$ is a term depending on the product of the Malliavin derivatives of $\lambda$ and $\sigma$ and higher-order Malliavin derivatives. Similarly, for the third term  we can see that 
\begin{equation}
\label{segona}
\begin{aligned}
0=&\E_t \left[ N^t_T\int_t^T\!\sigma_s\partial_x \cB (s,X_s,v_s)\sum_{i=1}^2 \int_s^T \!\!\!D_s^{B^i}\! \lambda_udu\, d\langle B,B^i\rangle_s \right.\\
-&\left.N^t_T\partial_x \cB (t,X_t,v_t)\int_t^T\!\! \sigma_s\sum_{i=1}^2 \int_s^T \!\!\!D_s^{B^i}\! \lambda_udu\, d\langle B,B^i\rangle_s\right]+T_5,
\end{aligned}
\end{equation}
where $T_5$ is again a term depending on the product of the Malliavin derivatives of $\lambda$ and $\sigma$ and higher-order Malliavin derivatives. Substituting in  \eqref{repr},  we have \begin{equation}
\begin{aligned}
\label{repr2}
\!\!\!CVA(t)=& \E_t\left[(1-N^t_T) \cB(t,X_t,v_t)\right]\\
+&\frac 12\E_t\left[(1-N^t_T)\left( \partial_{xxx}\!- \!\partial_{xx}\right)\cB(t,X_t,v_t)\!\!\int_t^T \!\! \sigma_s\sum_{i=1}^2\!\int_s^T\!\!\! D_s^{B^i}\! \sigma_u^2du\,d\langle B,B^i\rangle_s\right]\\
-&\E_t\left[N^t_T\partial_x \cB (t,X_t,v_t)\int_t^T\!\! 
\sigma_s)\sum_{i=1}^2 \int_s^T \!\!\!D_s^{B^i}\! \lambda_udu\, d\langle B,B^i\rangle_s\right ]+T_4+T_5\\
=& \E_t\left[(1-N^t_T) \cB(t,X_t,v_t)\right]+\frac 12\E_t\left[(1-N^t_T)\left( \partial_{xxx}\!- \!\partial_{xx}\right)\cB(t,X_t,v_t)\!\!\int_t^T d\langle X,M\rangle\right]\\
-&\E_t\left[N^t_T\partial_x \cB (t,X_t,v_t)\int_t^T\!\!  d\langle X,N\rangle_s\right ]+T_4+T_5\\
=&T_1'+T_2'+T_3'+T_4+T_5.
\end{aligned}
\end{equation}
Denoting by 
$\displaystyle
A_r^1:=E_r\left[\int_t^T\!\!\! \!d\langle M\! ,X\rangle_s\right ],
A_r^2:=E_r\left[\int_t^T\!\!\! \!d\langle N\! ,X\rangle_s\right ]
$
and recalling the expression of $v$, 
the above equality reads as
\begin{equation}
\begin{aligned}
\label{repr3}
\!\!\!CVA(t)=& \E_t\left[(1-N^t_T) \cB\left(t,X_t,\sqrt{\frac{M_T-\int_0^t\sigma_u^2du}{T-t}}\right)\right]\\
+&\frac 12\E_t\left[(1-N^t_T)\left( \partial_{xxx}\!- \!\partial_{xx}\right)\cB\left(t,X_t,\sqrt{\frac{M_T-\int_0^t\sigma_u^2du}{T-t}}\right)\!\!A_T^1\right]\\
-&\E_t\left[N^t_T\partial_x \cB \left(t,X_t,\sqrt{\frac{M_T-\int_0^t\sigma_u^2du}{T-t}}\right)A_T^2\right ]+T_4+T_5.
\end{aligned}
\end{equation}
The first three terms in the above equality are functions of the martingales $M,N, A^1$ and $A^2$ evaluated at time $T$. By a direct application of It\^o's formula, we finally rewrite \eqref{repr3} as 
\begin{equation}
\begin{aligned}
\label{repr4}
\!\!\!CVA(t)=& (1\!-\!N^t_t) \left[\cB\left(t,X_t,\sqrt{\frac{M_t\!-\!\int_0^t\sigma_u^2du}{T-t}}\right)+\frac 12\left( \partial_{xxx}\!- \!\partial_{xx}\right)\cB\left(t,X_t,\sqrt{\frac{M_t\!-\!\int_0^t\sigma_u^2du}{T-t}}\right)\!\!A_t^1\right]\\
-&N^t_t\partial_x \cB \left(t,X_t,\sqrt{\frac{M_t-\int_0^t\sigma_u^2du}{T-t}}\right)A_t^2
+T_1''+T_2''+T_3''+T_4+T_5\\
=&(1-N^t_t)\left [ \cB\left(t,X_t,\hat{v}_t\right)
+\frac 12\left( \partial_{xxx}\!- \!\partial_{xx}\right)\cB\left(t,X_t,\hat{v}_t\right)\!\!A_t^1\right ]
-N^t_t\partial_x \cB \left(t,X_t,\hat{v}_t\right)A_t^2\\
+&T_1''+T_2''+T_3''+T_4+T_5.
\end{aligned}
\end{equation}
All the terms  $T_1''+T_2''+T_3''$ can be expressed in terms of the quadratic variation and covariation of the martingales $N,M,A^1$, and $A^2$, which, in turn, may be expressed in terms of the Malliavin derivatives of $\sigma$ and $\lambda$ by using the Clark-Ocone-Haussman formula.

We may conclude that the error is controlled by the well known bounds on the Black \& Scholes pricing function and its derivatives, while those Malliavin derivatives, which become explicitly computable once the models are specified.
\section{Numerical results}

\label{sec5}

\begin{figure}[h]
\begin{center}
\includegraphics[scale=0.6]{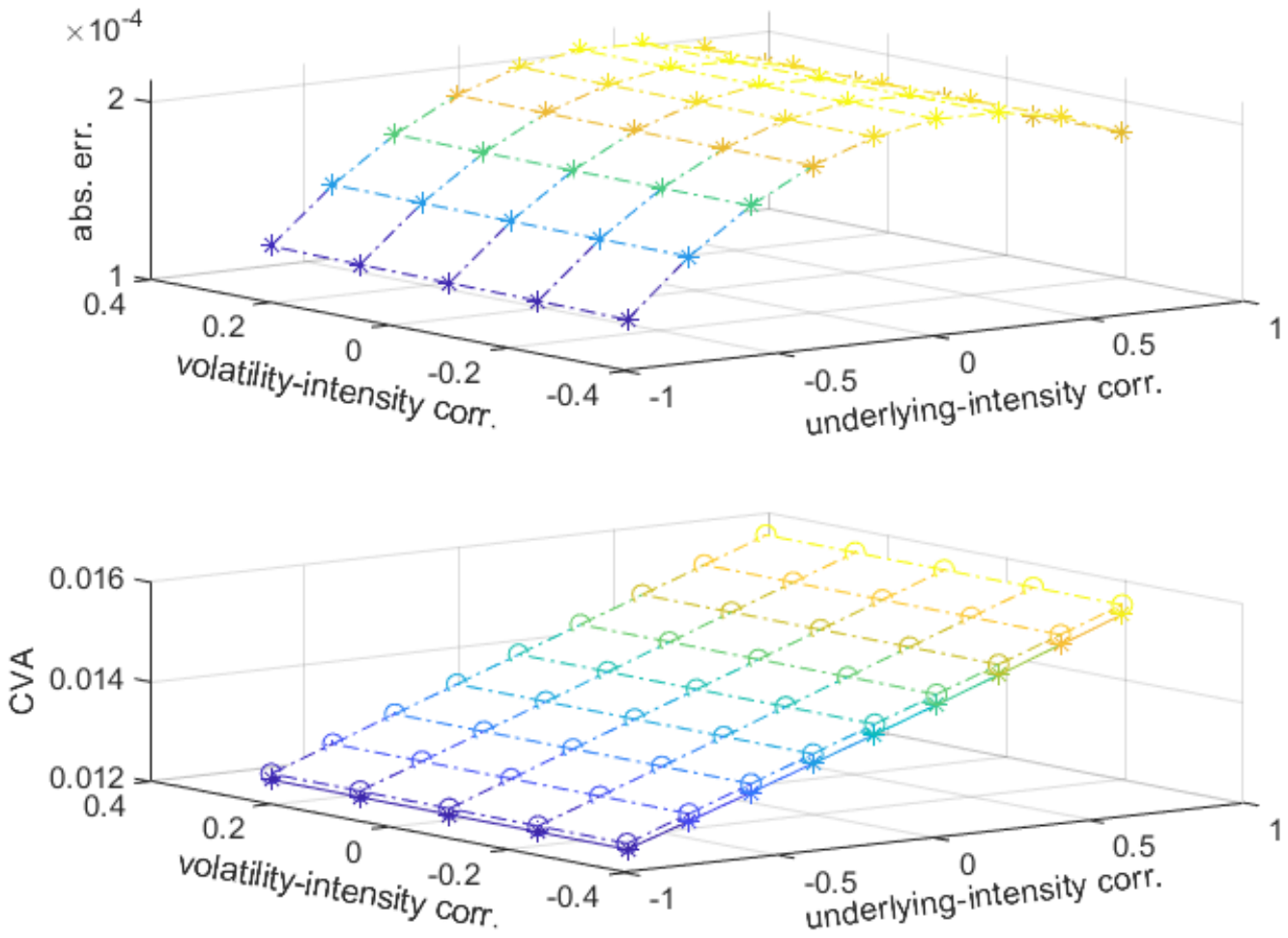}
\vspace{-5cm}
\caption{{\footnotesize CVA's $T=1/4$ - set A: Monte Carlo ($*$-) vs approximation (o-.).}}
\label{CVAfigT025_setA}
\end{center}
\end{figure}

\begin{figure}[h]
\begin{center}
\vspace{1cm}
\vspace{-7cm}
\includegraphics[scale=0.6]{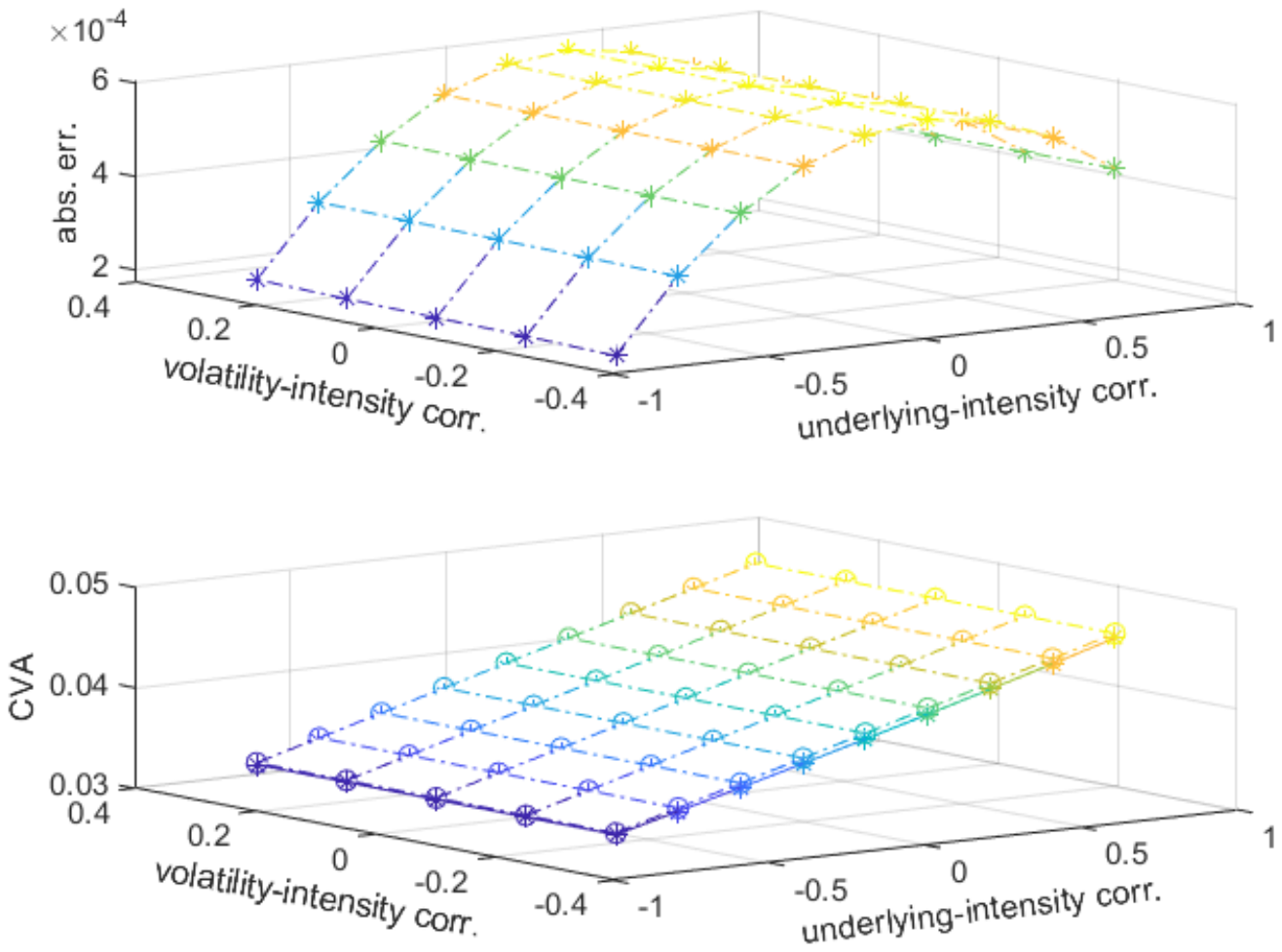}
\vspace{-5cm}
\caption{{\footnotesize CVA's $T=1/2$ - set A: Monte Carlo ($*$-) vs approximation (o-.).}}
\label{CVAfigT05_setA}
\end{center}
\end{figure}

\begin{figure}[h]
\begin{center}
\vspace{-4cm}
\vspace{-7cm}
\includegraphics[scale=0.6]{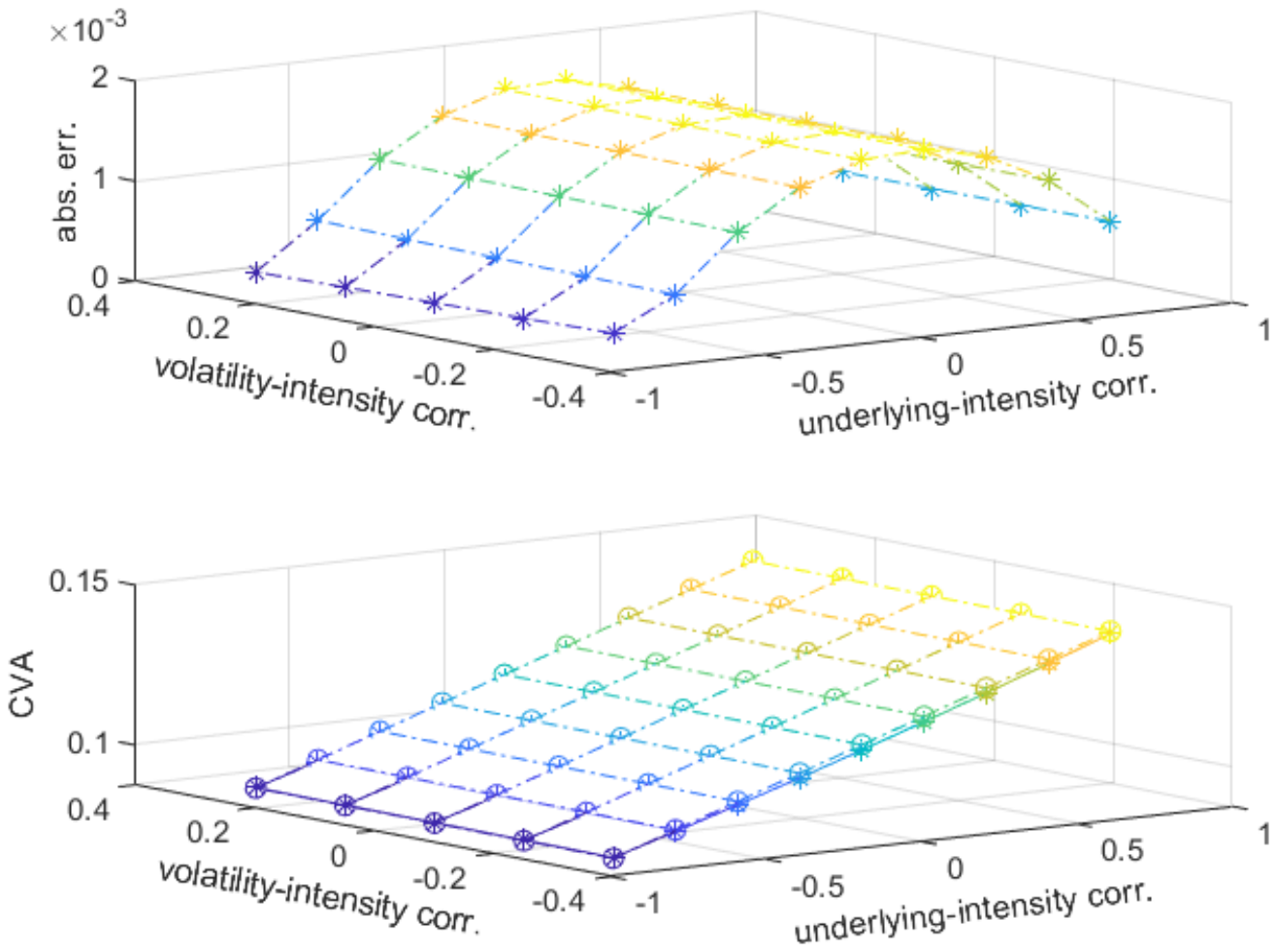}
\vspace{-5cm}
\caption{{\footnotesize CVA's $T=1$ - set A: Monte Carlo ($*$-) vs approximation (o-.).}}
\label{CVAfigT1_setA}
\end{center}
\end{figure}

\begin{figure}[h]
\begin{center}
\vspace{-5cm}
\includegraphics[scale=0.6]{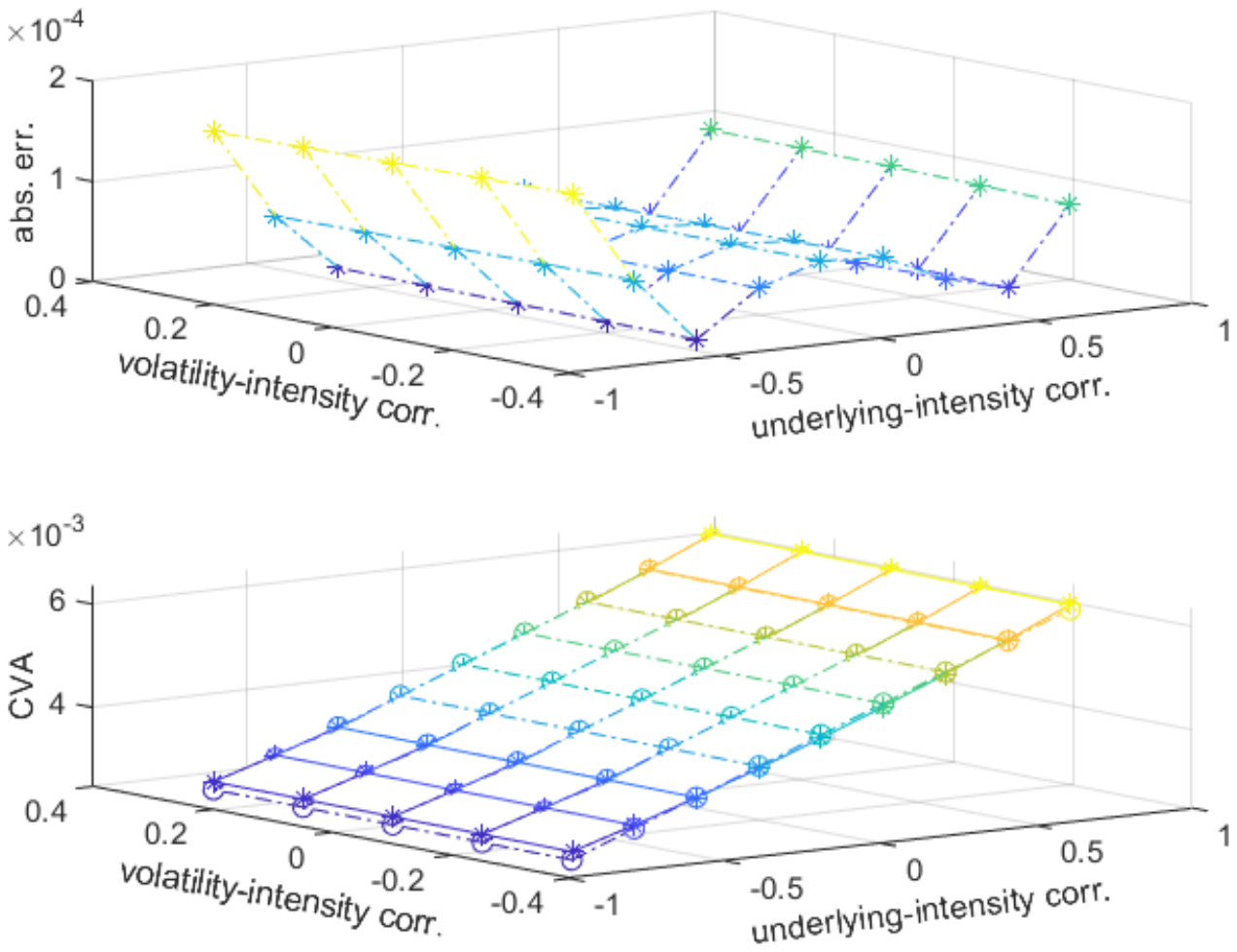}
\vspace{-5cm}
\caption{{\footnotesize CVA's $T=1/4$ - set B: Monte Carlo ($*$-) vs approximation (o-.).}}
\label{CVAfigT025_setB}
\end{center}
\end{figure}

\begin{figure}[h]
\begin{center}
\vspace{-6cm}
\includegraphics[scale=0.6]{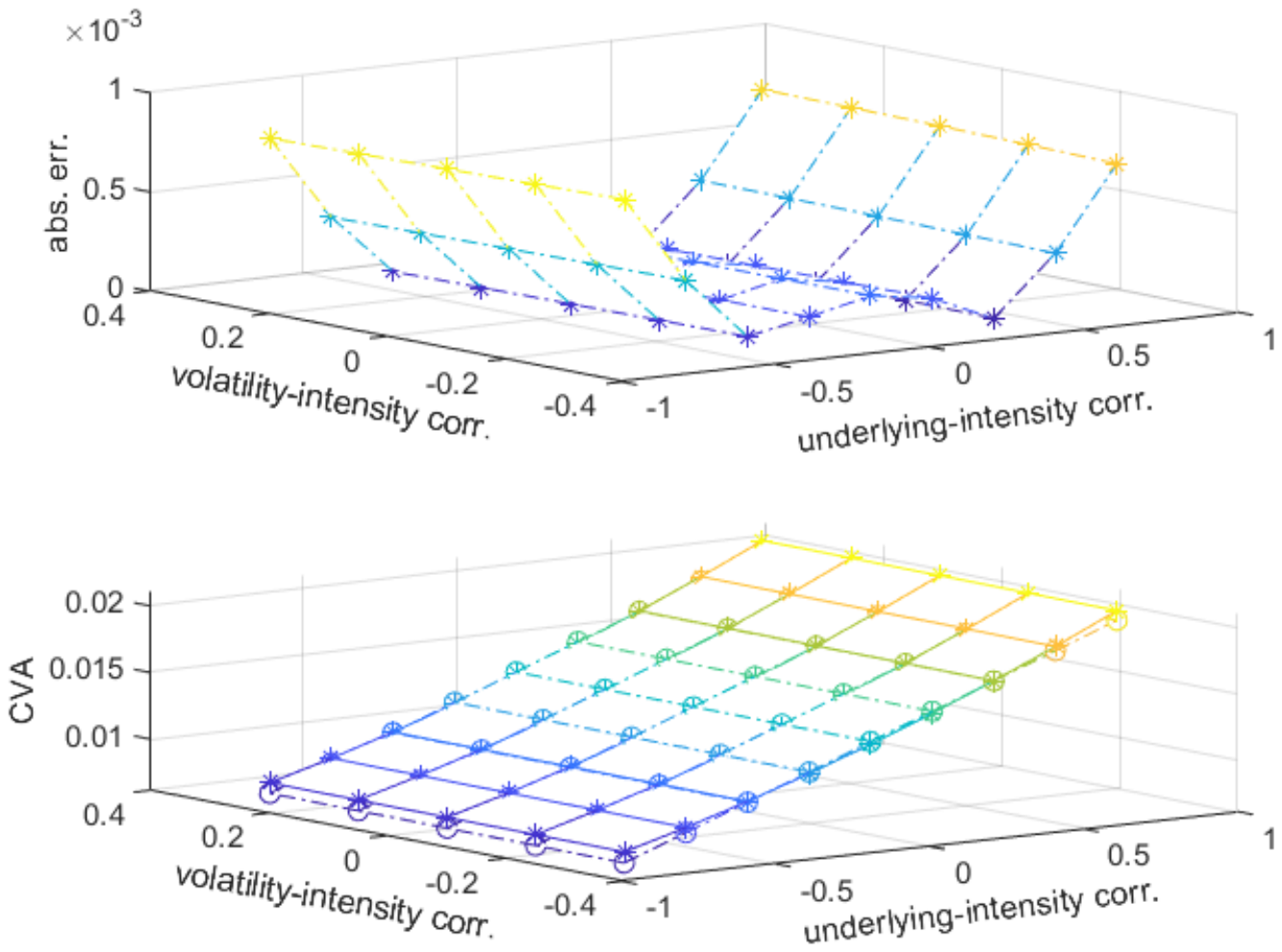}
\vspace{-5cm}
\caption{{\footnotesize CVA's $T=1/2$ - set B: Monte Carlo ($*$-) vs approximation (o-.).}}
\label{CVAfigT05_setB}
\end{center}
\end{figure}

\begin{figure}[h]
\begin{center}
\vspace{-7cm}
\includegraphics[scale=0.6]{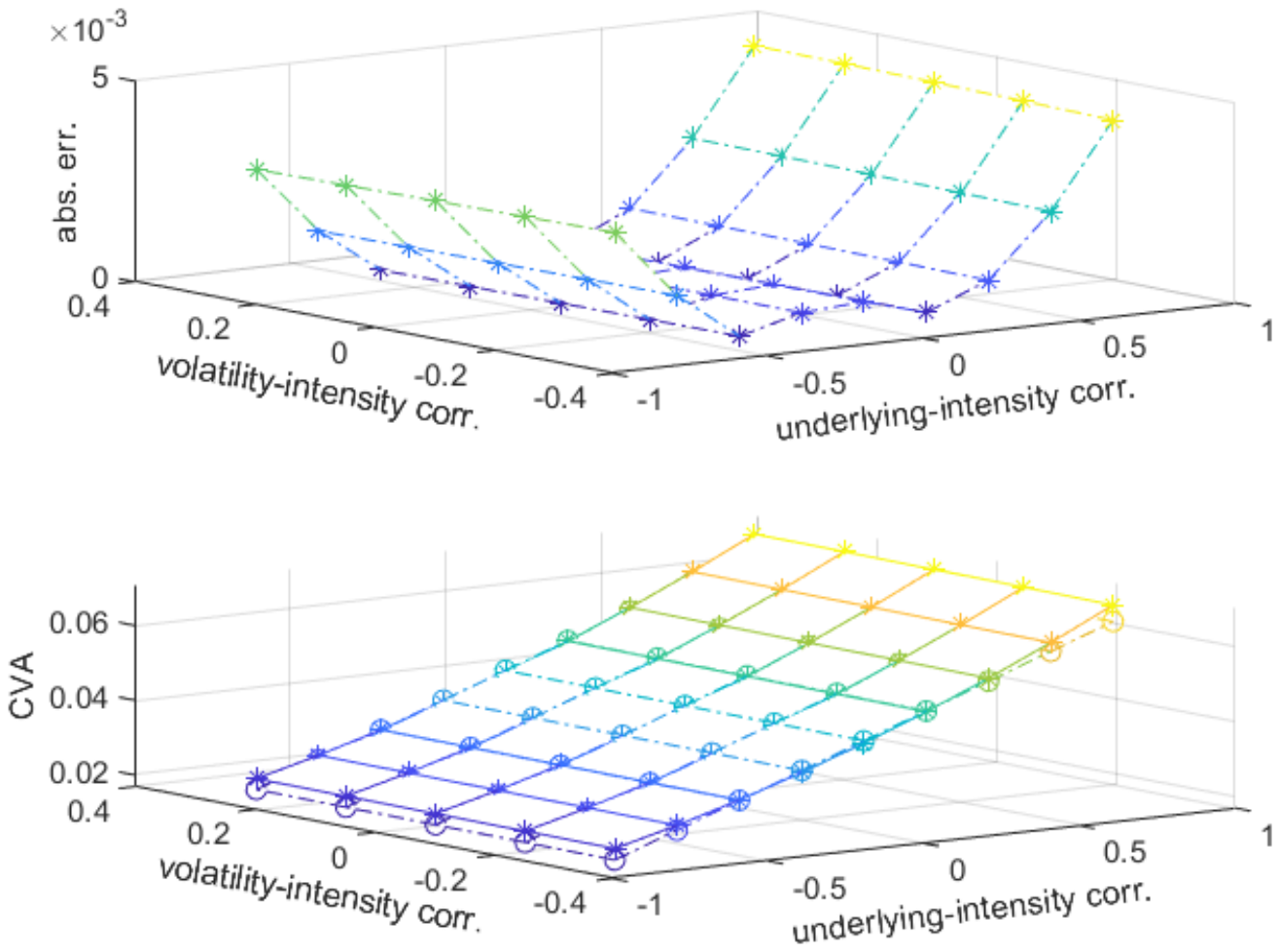}
\vspace{-5cm}
\caption{{\footnotesize CVA's $T=1$ - set B: Monte Carlo ($*$-) vs approximation (o-.).}}
\label{CVAfigT1_setB}
\end{center}
\end{figure}

In this section, we compare the numerical efficiency of our method with the standard benchmark Monte Carlo technique to evaluate (\ref{cva0}) in the rBergomi model. As said, we assume $r=0$ and $R=0$.  We simulated $N=10^6$ sample paths of the joint model rBergomi-CIR with a discretization of the time interval $[0,T]$ obtained by $100$ equi-spaced points. The exact covariance matrix of the three gaussian processes $(B, Z, W)$ has been used to get the sample paths (see \cite{AL21}). Hence, the rBergomi is simulated by standard multivariate normal generation, while for the intensity component we implemented the Euler discretization scheme with full truncation. We chose two sets of parameters for the intensity CIR process as in \cite{BV18}, set A and set B (see Table \ref{CIR_param}), which agree with those of calibrated default intensities. Then we fixed the rBergomi model parameters to $\nu=0.1$, $H=0.1$, $\sigma_0=0.08$ and the underlying-volatility correlation to $\eta=-0.2$,  and we considered a set of equi-spaced underlying-intensity and volatility-intensity correlations, $\rho$ and $\gamma$ respectively. In fact we would like to assess the quality of the our approximation which highlights the dependence  of the credit adjustment on the default intensity correlations, known as right/wrong-way-risk. All the computational procedures were implemented by using MatLab R2021b on a Intel(R) Core(TM) i7-4700MQ CPU @ 2.40GHz. The results of the approximations are reported in Tables (\ref{resT025_setA})-(\ref{resT1_setB}) and shown in the Figures (\ref{CVAfigT025_setA})-(\ref{CVAfigT1_setB}) for an ATM call option with $S_0=100$, $K=100$ and maturities $T=0.25, 0.5, 1$ years, respectively. The time to compute the set of $45$ adjustments with the Monte Carlo algorithm was approximately one hour, while about $30$ secs are needed for the computation of our approximation, which requires the numerical evaluation of multiple integrals. This step was done by using the MatLab build-in function.  We may notice that the approximation errors increase with maturity, ranging from $1.18e-04$ to $1.7e-03$ for set A and from $1.19e-06$ to $4.6e-03$ for set B. This was to be expected since in our approximation we freeze some terms at their initial value and therefore their contribution in  time dimension is lost.  We notice that the error due to the volatility-intensity correlation remains essentially constant across all simulations, while most of the variation is due to the underlying-intensity correlation, which proves to be the most relevant parameter. We notice the the errors pattern is different between the two intensity sets of parameters: for set A the larger errors come from small (underlying-intensity) correlations, almost the opposite for set B.  
 
\begin{figure}
\vspace{-5cm}
\begin{subfigure}{.5\textwidth}
  \centering
  \includegraphics[width=1.1\linewidth]{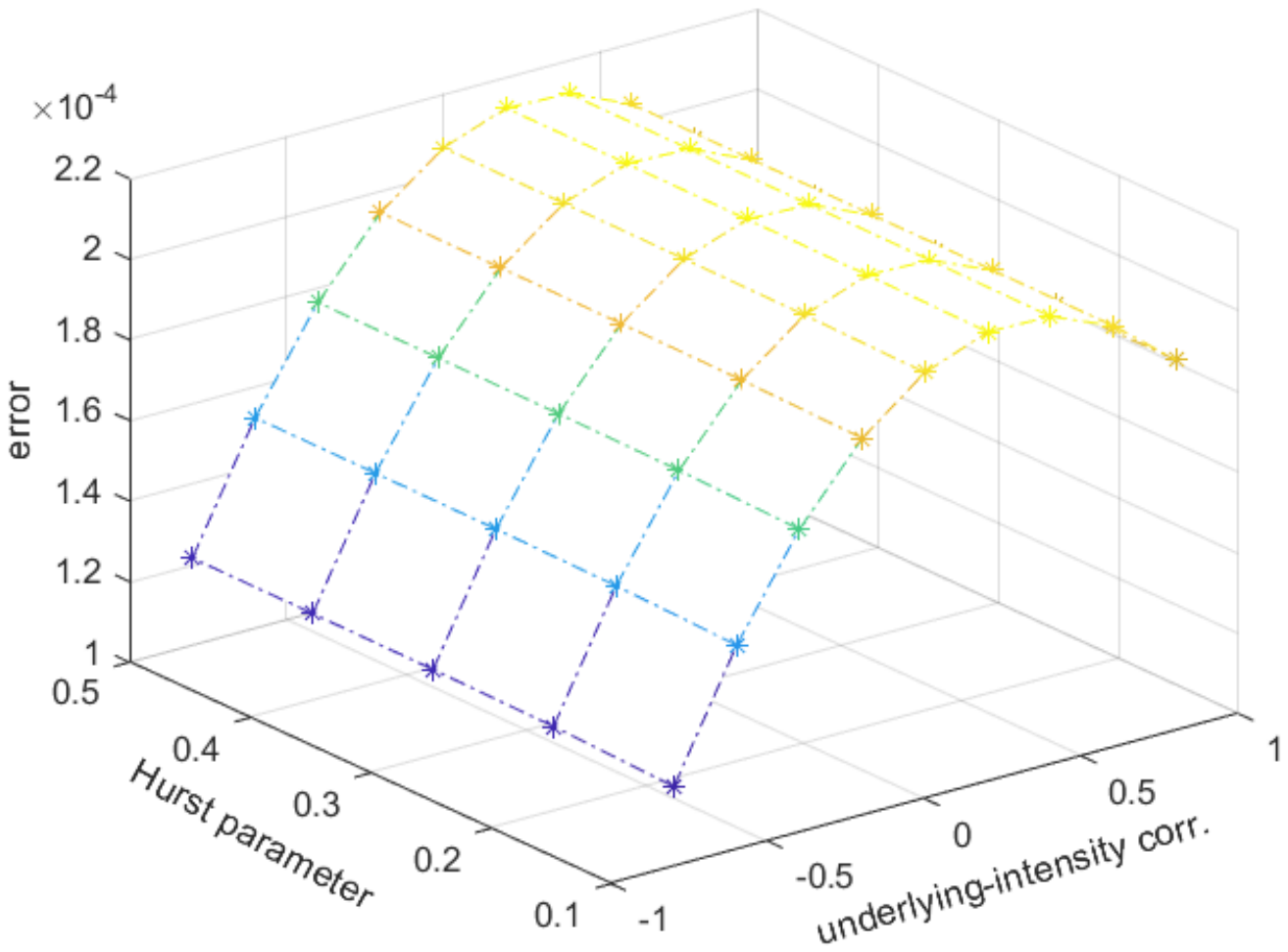}
  \vspace{-4cm}
  \caption{}
\end{subfigure}%
\begin{subfigure}{.5\textwidth}
  \centering
  \includegraphics[width=1.1\linewidth]{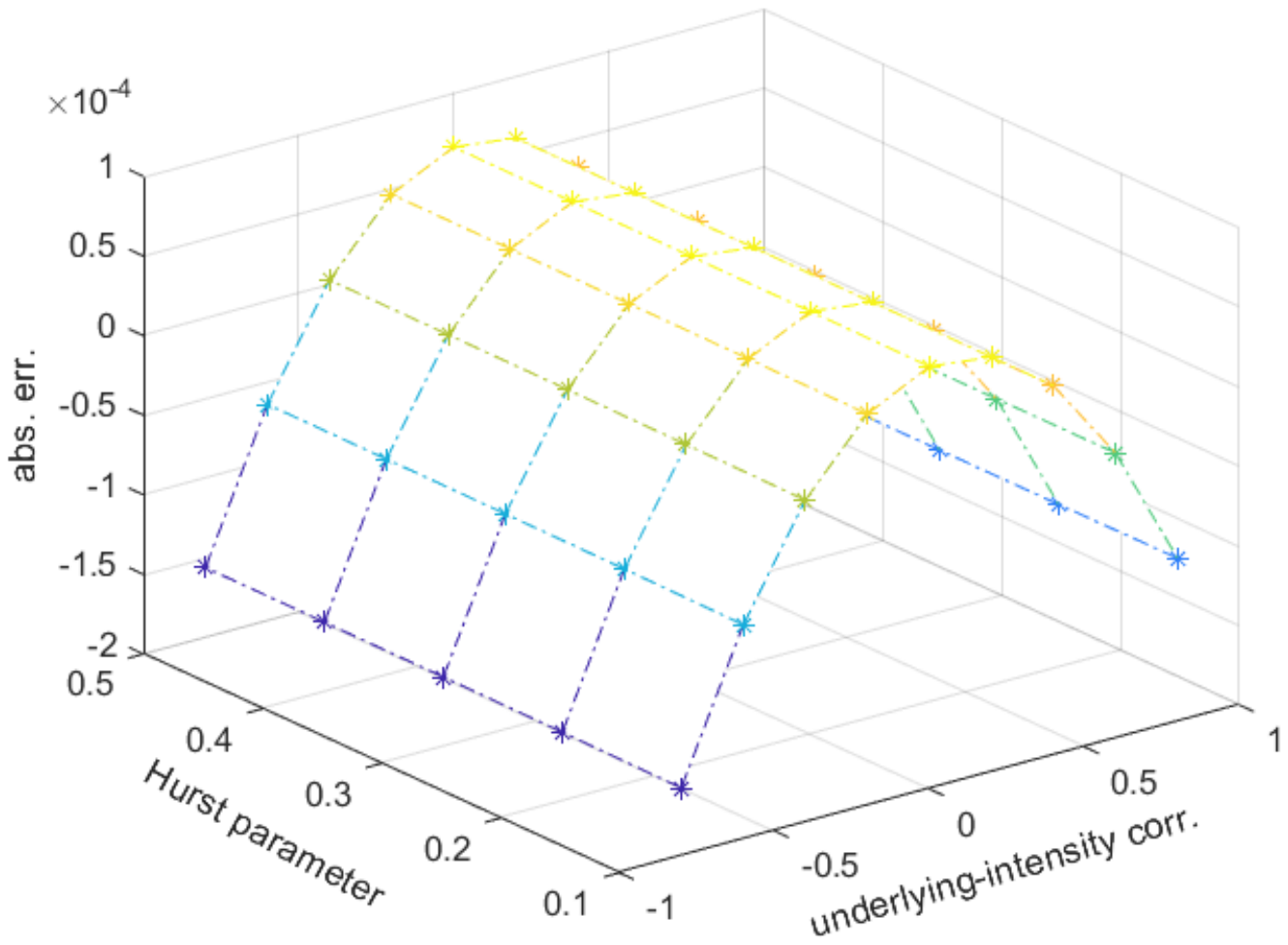}
  \vspace{-4cm}
  \caption{}
\end{subfigure}
\caption{{\footnotesize Approximation error as a function of the correlation $\rho$ and the Hurst parameter $H$ for sets A and B, $T=0.25$.}}
\label{cva_err_rH_T025}
\end{figure}

Finally, we considered also the sensitivity of our approximation to the Hurst parameter $H$. Figure (\ref {cva_err_rH_T025}) reports the errors between Monte Carlo estimate and the first order approximation as a function of the correlation $\rho$ and the Hurst parameter $H$. Even in this case, the dependence of the error from $H$ is almost constant and depends mainly on the correlation parameter, however remaning of order $10^{-4}$.

\begin{table}[t]
\centering
\small
\begin{tabular}{|c|c|c|c|c|}
 \hline
   & $\lambda_0$ & $q$  & $\mu$ &  $c$ \\ \hline
 Set A  & 0.035 & 0.35 & 0.035   & 0.1 \\
 Set B  & 0.01 & 0.8 & 0.02 & 0.2\\
  \hline
\end{tabular}
 \caption{ {\footnotesize Parameter sets for the CIR default intensity.}}
\label{CIR_param}
\end{table}

\newpage
\subsection{Results - Set A}

\begin{table}[h]
\centering
\small
\begin{tabular}{c|c|c|c|c|c}
\hline
$\rho \backslash \gamma$  &-0.3 &-0.15 &0  &0.15 & 0.3 \\ \hline
-0.8 &   1.186e-04  &  1.195e-04 &   1.199e-04 &   1.206e-04 &   1.214e-04\\
-0.6 &   1.501e-04  &  1.504e-04 &   1.508e-04 &   1.513e-04 &   1.520e-04\\
-0.4 &   1.748e-04  &  1.750e-04 &   1.753e-04 &   1.758e-04 &   1.764e-04\\
-0.2 &   1.933e-04  &  1.934e-04 &   1.937e-04 &   1.940e-04 &   1.945e-04\\
 0.0 &   2.056e-04  &  2.056e-04 &   2.057e-04 &   2.060e-04 &   2.064e-04\\
0.2  &   2.116e-04  &  2.116e-04 &   2.116e-04 &   2.118e-04 &   2.121e-04\\
0.4  &   2.115e-04  &  2.113e-04 &   2.112e-04 &   2.114e-04 &   2.116e-04\\
0.6  &   2.051e-04  &  2.048e-04 &   2.047e-04 &   2.048e-04 &   2.050e-04\\
0.8  &   1.926e-04  &  1.922e-04 &   1.920e-04 &   1.921e-04 &   1.920e-04\\
\hline
\end{tabular}
\caption{{\footnotesize Approximation errors, set A for CIR parameters. Here $T=1/4$, the risk-free call price  is $1.5877$. The lengths of the Monte Carlo confidence intervals for each price are in the range $(6.7651\mathrm{e}-05, 9.7582\mathrm{e}-05)$.}}
\label{resT025_setA}
\end{table}

\begin{table}[h]
\centering
\small
\begin{tabular}{c|c|c|c|c|c}
\hline
$\rho \backslash \gamma$  &-0.3 &-0.15 &0  &0.15 & 0.3 \\ \hline
-0.8 &   1.737e-04 &  1.783e-04 &  1.811e-04  &  1.845e-04 &  1.887e-04 \\
-0.6 &   3.288e-04 &  3.307e-04 &  3.330e-04  &  3.357e-04 &  3.391e-04\\
-0.4 &   4.479e-04 &  4.492e-04 &  4.509e-04  &  4.531e-04 &  4.559e-04\\
-0.2 &   5.332e-04 &  5.339e-04 &   5.350e-04 &  5.366e-04 &  5.388e-04\\
0.0  &   5.846e-04 &  5.847e-04 &   5.853e-04 &  5.864e-04 &  5.881e-04\\
0.2  &   6.022e-04 &  6.017e-04 &   6.018e-04 &  6.024e-04 &  6.037e-04\\
0.4  &   5.861e-04 &  5.851e-04 &   5.847e-04 &  5.849e-04 &  5.858e-04\\
0.6  &   5.365e-04 &  5.349e-04 &   5.340e-04 &  5.339e-04 &  5.347e-04\\
0.8  &   4.536e-04 &  4.513e-04 &   4.501e-04 &  4.500e-04 &  4.493e-04\\
\hline
\end{tabular}
\caption{{\footnotesize Approximation errors, set A for CIR parameters. Here $T=1/2$, the risk-free call price  is $2.2450$. The lengths of the Monte Carlo confidence intervals for each price are in the range $(1.8252\mathrm{e}-04, 3.0014\mathrm{e}-04)$.}}
\label{resT05_setA}
\end{table}

\begin{table}[h]
\centering
\small
\begin{tabular}{c|c|c|c|c|c}
\hline
$\rho \backslash \gamma$  &-0.3 &-0.15 &0  &0.15 & 0.3 \\ \hline
-0.8 & 2.1015e-04 &  1.8692e-04 &  1.7121e-04 &  1.5363e-04 &  1.3338e-04 \\
-0.6 & 5.2435e-04 &  5.3558e-04 &  5.4804e-04 &  5.6225e-04 &  5.7867e-04\\
-0.4 & 1.0766e-03 &  1.0844e-03 &  1.0937e-03 &  1.1047e-03 &  1.1179e-03 \\
-0.2 & 1.4554e-03 &  1.4600e-03 &  1.4662e-03 &  1.4743e-03 &  1.4845e-03 \\
0.0  & 1.6613e-03 &  1.6627e-03 &  1.6660e-03 &  1.6712e-03 &  1.6787e-03\\
 0.2 & 1.6947e-03 &  1.6930e-03 &  1.6934e-03 &  1.6960e-03 &  1.7011e-03\\
0.4  & 1.5560e-03 &  1.5513e-03 &  1.5489e-03 &  1.5491e-03 &  1.5523e-03\\
0.6  & 1.2457e-03 &  1.2379e-03 &  1.2331e-03 &  1.2313e-03 &  1.2332e-03\\
 0.8 & 7.6526e-04 &  7.5409e-04 &  7.4722e-04 &  7.4475e-04 &  7.4117e-04\\  \hline

\end{tabular}
\caption{{\footnotesize Approximation errors, set A for CIR parameters. Here $T=1$, the risk-free call price  is  $3.1742$. The lengths of the Monte Carlo confidence intervals for each price are in the range $(4.9156\mathrm{e}-04, 9.4118\mathrm{e}-04)$.}}
\label{resT1_setA}
\end{table}

\clearpage
\subsection{Results - Set B}

\begin{table}[h]
\centering
\small
\begin{tabular}{c|c|c|c|c|c}
\hline
$\rho \backslash \gamma$  &-0.3 &-0.15 &0  &0.15 & 0.3 \\ \hline
-0.8 &  1.6115e-04 &  1.5887e-04 &  1.5728e-04 &  1.5560e-04 &  1.5378e-04\\
-0.6 &  6.7428e-05 &  6.6354e-05 &  6.5250e-05 &  6.4024e-05 &  6.2642e-05\\
-0.4 &  1.1889e-06 &  1.7841e-06 &  2.4470e-06 &  3.2382e-06 &  4.2046e-06\\
-0.2 &  4.5520e-05 &  4.5687e-05 &  4.5942e-05 &  4.6338e-05 &  4.6921e-05\\
 0.0 &  6.5712e-05 &  6.5452e-05 &  6.5321e-05 &  6.5344e-05 &  6.5567e-05\\
0.2  &  6.1873e-05 &  6.1201e-05 &  6.0678e-05 &  6.0338e-05 &  6.0237e-05\\
0.4  &  3.4001e-05 &  3.2921e-05 &  3.2032e-05 &  3.1366e-05 &  3.0991e-05\\
0.6  &  1.7912e-05 &  1.9391e-05 &  2.0617e-05 &  2.1556e-05 &  2.2129e-05\\
0.8  &  9.3819e-05 &  9.5716e-05 &  9.7231e-05 &  9.8323e-05 &  9.9291e-05\\
\hline
\end{tabular}
\caption{{\footnotesize Approximation errors, set B for CIR parameters. Here $T=1/4$, the risk-free call price  is  $1.5877$. The lengths of the Monte Carlo confidence intervals for each price are in the range $(1.4796e-05, 4.6716e-05)$.}}
\label{resT025_setB}
\end{table}

\begin{table}[h]
\centering
\small
\begin{tabular}{c|c|c|c|c|c}
\hline
$\rho \backslash \gamma$  &-0.3 &-0.15 &0  &0.15 & 0.3 \\ \hline
-0.8 & 8.2741e-04 &  8.1698e-04 &  8.0882e-04 &  8.0048e-04 &  7.9149e-04\\
-0.6 & 3.8350e-04 &  3.7768e-04 &  3.7173e-04 &  3.6546e-04 &  3.5854e-04\\
-0.4 & 6.5879e-05 &  6.2609e-05 &  5.8946e-05 &  5.4764e-05 &  4.9882e-05\\
-0.2 & 1.2894e-04 &  1.2985e-04 &  1.3128e-04 &  1.3333e-04 &  1.3618e-04\\
0.0  & 2.0179e-04 &  2.0055e-04 &  1.9992e-04 &  1.9991e-04 &  2.0072e-04\\
0.2  & 1.5314e-04 &  1.4983e-04 &  1.4707e-04 &  1.4509e-04 &  1.4416e-04\\
0.4  & 1.7027e-05 &  2.2576e-05 &  2.7401e-05 &  3.1240e-05 &  3.3776e-05\\
0.6  & 3.0968e-04 &  3.1744e-04 &  3.2417e-04 &  3.2963e-04 &  3.3338e-04\\
0.8  & 7.2598e-04 &  7.3603e-04 &  7.4441e-04 &  7.5091e-04 &  7.5625e-04\\
\hline
\end{tabular}
\caption{{\footnotesize Approximation errors, set B for CIR parameters. Here $T=1/2$ and the price of the risk-free call is $2.2450$. The lengths of the Monte Carlo confidence intervals for each price are in the range $(4.0954\mathrm{e}-05,1.6693\mathrm{e}-04)$.}}
\label{resT05_setB}
\end{table}

\begin{table}[h]
\centering
\small
\begin{tabular}{c|c|c|c|c|c}
\hline
$\rho \backslash \gamma$  &-0.3 &-0.15 &0  &0.15 & 0.3 \\ \hline
-0.8 & 3.0554e-03 &  3.0129e-03 &  2.9782e-03 &  2.9431e-03 &  2.9057e-03 \\
-0.6 & 1.3068e-03 &  1.2805e-03 &  1.2548e-03 &  1.2278e-03 &  1.1989e-03\\
-0.4 & 1.1803e-04 &  1.0141e-04 &  8.4800e-05 &  6.7575e-05 &  4.7144e-05\\
-0.2 & 5.2478e-04 &  5.3076e-04 &  5.3764e-04 &  5.4572e-04 &  5.5712e-04\\
0.0  & 6.2528e-04 &  6.2068e-04 &  6.1724e-04 &  6.1659e-04 &  6.1960e-04\\
0.2  & 1.8308e-04 &  1.6810e-04 &  1.5603e-04 &  1.4691e-04 &  1.4118e-04\\
0.4  & 8.0508e-04 &  8.2955e-04 &  8.5101e-04 &  8.6852e-04 &  8.8145e-04\\
0.6  & 2.3448e-03 &  2.3792e-03 &  2.4097e-03 &  2.4354e-03 &  2.4540e-03\\
0.8  & 4.4474e-03 &  4.4929e-03 &  4.5321e-03 &  4.5636e-03 &  4.5870e-03\\
\hline
\end{tabular}
\caption{{\footnotesize Approximation errors, set B for CIR parameters. Here $T=1$ and the price of the risk-free call is $2.2450$. The lengths of the Monte Carlo confidence intervals for each price are in the range $(1.2015e-04,6.0014e-04)$.}}
\label{resT1_setB}
\end{table}

\appendix\section {Malliavin Calculus}

Here we introduce the main concepts of Malliavin Calculus used in this work. We refer to \cite{Nu} for a deeper approach to this field.

Let ${\cal S}$ be the set of random variables of the form
\begin{equation}
\label{eq:2.1}
F=f(W(h_{1}),\ldots ,W(h_{n})),  
\end{equation}
with $h_{1},\ldots ,h_{n}\in L^2([0,T])$, $W(h_i)$ denotes the Wiener integral of the function $h_i$, for $i=1,..,n$, and $f\in C_{b}^{\infty }(\mathbb{R}^n) $ 
(i.e., $f$ and all its partial derivatives are bounded). The Malliavin 
derivative of $F$, $D F$,  is defined
as the stochastic process in  $L^{2}(\Omega \times [0,T])$ given by 
\begin{equation*}
D_{s}F=\sum_{j=1}^{n}{\frac{\partial f}{\partial x_{j}}}(W(h_{1}),\ldots ,W(h_{n}))(s)h_j(s).
\end{equation*}
We can also define the iterated Malliavin derivative operator as
\begin{equation*}
D_{s_{1},\ldots ,s_{m}}^{m}F=D_{s_{1}}\ldots D_{s_{m}}F,\qquad s_{1},\ldots
,s_{m}\in [0,T].
\end{equation*}
These operators are closable  in $
L^{p}(\Omega )$ for any $p\geq 1$ and we denote by ${\mathbb{D}}^{n,p}$ the
closure of ${\cal S}$ with respect to the norm
$$
||F||_{n,p}=\left( E\left| F\right|
^{p}+\sum_{i=1}^{n}E||D^{i}F||_{L^{2}([0,T]^{i})}^{p}\right) ^{\frac{1}{p}}.
$$
We will also make use of the notation $\mathbb{L}^{n,p}=\mathbb{D}_{W}^{n,p}(L^2([0,T])).$ 

The adjoint of the derivative operator $D$ is the divergence 
operator $\delta $, that coincides with the Skorohod integral. That is, the domain of 
$\delta $, denoted by Dom $\delta $, is the set of processes 
$u\in L^{2}(\Omega \times [0,T])$ such that there exists 
$\delta (u)\in L^{2}(\Omega )$ such that
\begin{equation}\label{eq:ibpf}
E(\delta (u)F)=E\left(\int_{0}^{T}(D_{s}F)u_{s}ds\right),\qquad \hbox{\rm for every}
\;\;F\in {\cal S}.  
\end{equation}
We use the notation $\delta (u)=\int_{0}^{T}u_{s}dW_{s}$.  It is
known that $\delta $ is an extension of the classical It\^o integral in the sense that $\delta$, applied to adapted and square integrable processes,  coincides with the classical It\^o  integral.  We also know that the space 
${\mathbb{L}}^{1,2}$ is included in the domain of $\delta$. 

\vspace{0.2cm}

The following result gives us an explicit martingale representation for random variables in  $\mathbb{D}_{W}^{1,2}$.
\begin{theorem}[The Clark-Ocone formula]
\label{CO}
Consider a random variable $F\in \mathbb{D}_{W}^{1,2}$.Then
$$
F=E(F)+\int_0^T E_r(D_r F)dW_r.
$$
\end{theorem}

Finally, we recall the anticipating It\^{o}’s formula
\begin{proposition}
\label{Ito}
Assume the model (\ref{SDEsystem0}) where $\sigma\in \mathbb{L}^{1,2}$ and consider the processes $Y,Z$ defined $Y_t=\int_t^T \sigma^2_{s}ds$. Let
$F:[0,T]\times \mathbb{R}^{3}\rightarrow \mathbb{R}$ be a function
in $C^{1,2} ([0,T]\times \mathbb{R}^{2})$ such that there exists a
positive constant $C$ such that, for all $t\in \left[ 0,T\right]
,$ $F$ and its partial derivatives evaluated in $\left(
t,X_{t},Y_{t},\right)$ are bounded by $C.$ Then it follows that
\begin{eqnarray}
F(t,X_{t},Y_{t}) &=&F(0,X_{0},Y_{0})+\int_{0}^{t}{\partial _{s}F}%
(s,X_{s},Y_{s})ds \nonumber\\
&&-\frac12\int_{0}^{t}{\partial _{x}F}(s,X_{s},Y_{s})\sigma^2_sds\nonumber\\
&&+\int_{0}^{t}{\partial _{x}F}(s,X_{s},Y_{s})\sigma_{s} dW_{s} \nonumber\\
&&-\int_{0}^{t}{\partial _{y}F}(s,X_{s},Y_{s})\sigma^2_{s}+
\int_{0}^{t}{\partial _{xy}^{2}F}(s,X_{s},Y_{s})\Theta _{s}ds \nonumber\\
&&-\int_{0}^{t}{\partial _{z}F}(s,X_{s},Y_{s})a^2_{s} \nonumber\\
&&+\frac{1}{2}\int_{0}^{t}{\partial
_{xx}^{2}F}(s,X_{s},Y_{s},Z_s)\sigma^2_{s}ds 
\label{aito}
\end{eqnarray}
where $\Theta _{s}:=(\int_{s}^{T}D_{s}\sigma_{r}^{2}dr)\sigma _{s}$.
\end{proposition}

\end{document}